\documentclass[11pt,twoside]{article}
\usepackage[english]{babel}
\usepackage{times,subeqnarray}
\usepackage{url}
\newif\myifpdf
\ifx\pdfoutput\undefined
   \usepackage[dvips]{graphicx}
\else
   \pdfoutput=1        
   \usepackage[pdftex]{graphicx}
\fi
\usepackage{apatitlepages}
\usepackage{setspace} 
\usepackage{psydraft}
\usepackage{apa}       

\newcommand{\wij}{w_{ij}}

\newcommand{\oneo}[1]{\frac{1}{#1}}

\myifpdf
  \DeclareGraphicsExtensions{.pdf,.eps,.png,.jpg,.mps,.tif}
\fi

\def\myheading{ Learning Through Time }

\begin{document}
\bibliographystyle{apa}

\sloppy
\raggedbottom

\def\mytitle{ Learning Through Time in the Thalamocortical Loops }

\def\myauthor{Randall C. O'Reilly, Dean Wyatte, and John Rohrlich\\
  Department of Psychology and Neuroscience \\
  University of Colorado Boulder \\
  345 UCB\\
  Boulder, CO 80309\\
  {\small randy.oreilly@colorado.edu}\\}

\def\mynote{arXiv.org q-bio.NC preprint version of submitted manuscript}

\def\myabstract{
  We present a comprehensive, novel framework for understanding how the neocortex, including the thalamocortical loops through the deep layers, can support a temporal context representation in the service of predictive learning.  Many have argued that predictive learning provides a compelling, powerful source of learning signals to drive the development of human intelligence: if we constantly predict what will happen next, and learn based on the discrepancies from our predictions (error-driven learning), then we can learn to improve our predictions by developing internal representations that capture the regularities of the environment (e.g., physical laws governing the time-evolution of object motions).  Our version of this idea builds upon existing work with simple recurrent networks (SRN's), which have a discretely-updated temporal context representations that are a direct copy of the prior internal state representation.  We argue that this discretization of temporal context updating has a number of important computational and functional advantages, and further show how the strong alpha-frequency (10hz, 100ms cycle time) oscillations in the posterior neocortex could reflect this temporal context updating.  Specifically, layer 5b intrinsically bursting neurons fire at the alpha frequency, and trigger an updating of the layer 6 regular spiking neurons that project down to the thalamus, and from there go back up to layer 4 and layer 6 -- this thalamocortical loop sustains the temporal context representation as the system develops a prediction about what will happen next.  When {\em next} inevitably happens, any resulting discrepancy can drive a biologically-based form of error-driven learning, which we have developed as part of the Leabra framework.  Thus, we refer to this new model as LeabraTI (temporal integration).  We examine a wide range of data from biology to behavior through the lens of this LeabraTI model, and find that it provides a unified account of a number of otherwise disconnected findings, all of which converge to support this new model of neocortical learning and processing.  We describe an implemented model showing how predictive learning of tumbling object trajectories can facilitate object recognition with cluttered backgrounds.
}

\titlesepage{\mytitle}{\myauthor}{\mynote}{\myabstract}





\pagestyle{myheadings}

How does the neocortex support the remarkable learning abilities that enable humans (and other mammals) to acquire the vast majority of our intelligence?  In this paper, we advance a comprehensive new framework for neocortical learning that leverages the biological properties of the thalamocortical loops (which are ubiquitous throughout the extent of the neocortex) to support {\em predictive learning}: learning from the differences between {\em expectations} versus actual {\em outcomes}.  This framework builds upon our existing neocortical learning framework, {\em Leabra} \cite{OReilly96,OReillyMunakata00,OReillyMunakataFrankEtAl12}, and adds a temporal integration mechanism, so we refer to this new framework as {\em LeabraTI}.  This temporal integration property, which depends on several features of the thalamocortical loops, enables the network to maintain prior temporal context in a way that can be leveraged to form increasingly accurate predictions or expectations about what will happen next.  This framework requires processing to be discretized over time, to distinguish between expectations versus outcomes as two sequential states of activation, and we find that this fits surprisingly well with the increasing evidence of a strong alpha frequency (10 Hz) modulation of neocortical processing \cite{LorinczKekesiJuhaszEtAl09,FranceschettiGuatteoPanzicaEtAl95,BuffaloFriesLandmanEtAl11,LuczakBarthoHarris13}, and corresponding behavioral data suggesting that perception is also discretized (at least to some extent) at the alpha frequency \cite{VanRullenKoch03b}.  Overall, we find that this learning framework integrates a wide range of previously unconnected biological and behavioral data, under a coherent, computationally powerful model.  In addition to reviewing and synthesizing this diverse body of empirical data, we demonstrate the computational power of this model in the context of object recognition with cluttered visual displays.

We begin with a brief review of the major threads of thought about how neocortical learning functions, at both the biological and functional / computational levels, to properly situate the LeabraTI model.  The quest to understand the essential form of human cognitive learning has a long history, including the classical contributions of \incite{James90} and \incite{Hebb49}, which emphasized what we now call the {\em Hebbian} correlational learning principle: ``neurons that fire together, wire together.''  At a functional level, Hebbian learning is thought to drive the encoding of important statistical structure in the external world (e.g., in an {\em internal model} of the environment), which then enables more sophisticated forms of inference or reasoning about the world.  Much of the recent focus on basic mechanisms of neural plasticity at the synaptic level has remained focused on the Hebbian paradigm, including considerable excitement about spike-timing dependent plasticity (STDP) \cite{BiPoo98,MarkramLubkeSakmann97}.  However, STDP is increasingly being recognized as essentially a product of the artificial conditions used to invoke it \cite{ShouvalWangWittenberg10}, and with realistic spike trains it reduces to a more classical form of Hebbian learning \cite{SongMillerAbbott00,ShouvalWangWittenberg10}.  In any case, there are very few, if any, computational models relying strictly on any form of Hebbian learning that achieve the signature capabilities of the mammalian neocortex (e.g., high-performance visual object recognition).  This is because Hebbian learning relies on strictly local, correlational signals that, while generally useful, are not actually sufficient to converge on the more complex representations that are often needed for real-world problems \cite{OReillyMunakata00,OReillyMunakataFrankEtAl12}.

Instead, most of the computationally-motivated work on high-performance learning algorithms has centered on error-driven learning mechanisms, most prominently the venerable error backpropagation algorithm \cite{RumelhartHintonWilliams86b}, and the support vector machine (SVM) \cite{CortesVapnik95}.  Error backpropagation for example is mathematically derived to solve whatever problem is posed to it, through the incremental process of error minimization, and has proven its value in a large number of models.  There has been a recent resurgence of interest in error backpropagation, which plays a central role in recent {\em deep learning} networks \cite{CiresanMeierGambardellaEtAl10,CiresanMeierSchmidhuber12,KrizhevskySutskeverHinton12,BengioCourvilleVincent13}.  Indeed, the best performance on a range of different benchmark tasks has been achieved with purely backpropagation networks, combined with a number of modern optimizations and tricks \cite{CiresanMeierSchmidhuber12,KrizhevskySutskeverHinton12,BengioCourvilleVincent13}.

Error-driven learning mechanisms can also achieve, more effectively, the same computational goal as Hebbian learning: developing an internal model of the structure of the external world.  Early work focused on the autoencoder model, where a network learns to reconstruct the inputs it receives, typically through a more compressed internal representation \cite{Pollack90,RumelhartMcClelland86}, and this is how backpropagation learning is often used in deep learning networks \cite{BengioCourvilleVincent13}.  More recently, the Bayesian framework has been leveraged for this same purpose, to create a {\em generative model} of the environment \cite{DayanHintonNealEtAl95,Friston05,Friston10}.  This framework can be traced all the way back to the notion of {\em recognition by synthesis} advanced by Helmholtz, and has been exploited in other neural learning frameworks as well \cite[e.g.,]{RaoBallard99,CarpenterGrossberg87,HintonSalakhutdinov06}.

In this paper, we focus on a specific form of the reconstructive or generative learning idea, which we call {\em predictive learning}, where instead of just reproducing or generating the current inputs, the network learns to predict what will happen next.  This idea was pioneered initially in the backpropagation framework, using simple recurrent networks (SRNs) \cite{Elman90,Elman91,Jordan89}, and has been leveraged in other frameworks as well \cite{SchusterPaliwal97,HawkinsBlakeslee04,GeorgeHawkins09}.  Generative and predictive forms of learning are particularly compelling because they provide a ubiquitous source of learning signals: if you attempt to predict everything that happens next, then every single moment is a learning opportunity.  This kind of pervasive learning can for example explain how an infant seems to magically acquire such a sophisticated understanding of the world, despite their seemingly inert overt behavior \cite{ElmanBatesJohnsonEtAl96} --- they are becoming increasingly expert predictors of what they will see next, and as a result, developing increasingly sophisticated internal models of the world.  In the context of language, predictive learning can drive the induction of sophisticated internal representations of syntactic categories, for example \cite{Elman91}.

We can now situate the LeabraTI model in the above context.  The Leabra framework on which LeabraTI is based is founded on a biologically-plausible form of error backpropagation \cite{OReilly96}, combined with a Hebbian associative learning mechanism (Leabra stands for Local, Error-driven and Associative, Biologically Realistic Algorithm --- it is pronounced like ``Libra'' and is intended to connote the {\em balance} of different factors).  A recent revision \cite{OReillyMunakataFrankEtAl12} includes a unified error-driven and Hebbian learning mechanism derived directly from a highly detailed biophysical model of STDP \cite{UrakuboHondaFroemkeEtAl08}.  The core idea for the error-driven learning aspect is simply that we learn when our expectations are violated.  This implies that we are constantly forming such expectations, so that violations thereof will provide valuable learning signals.  Specifically, learning in Leabra is driven by the difference between two subsequent states of activations in the network: the expectation or {\em minus phase}, compared with the outcome or {\em plus phase}.  This use of a temporal difference to drive learning is borrowed from the original Boltzmann machine and variants \cite{AckleyHintonSejnowski85,HintonMcClelland88a,GallandHinton90}, and contrasts with other frameworks that use the difference between top-down and bottom-up signals to drive learning \cite{Friston05,Friston10,GeorgeHawkins09,RumelhartHintonWilliams86b}.  Computationally, the LeabraTI framework just adds one key additional mechanism to standard Leabra: the ability to sustain the temporal context information necessary for generating predictions based on prior inputs and states of the network.  In this respect, it is very similar to the SRN modification of error backpropagation, and indeed there is a direct mathematical relationship between LeabraTI and the SRN, as we explain below.  However, LeabraTI generalizes the role of temporal context representations in important ways beyond the SRN framework, and provides a detailed biological account for how such context representations can be implemented in the neocortex.

In the remainder of this paper, we introduce the specific ideas for how the thalamocortical and neocortical laminar structure supports temporal integration and learning, and then review the relevant empirical literature across the biology and behavioral domains that bears on the specific computationally and biologically-motivated claims of the LeabraTI model.  Then, we present an application of the model to object recognition in cluttered visual scenes, and conclude with a discussion including comparisons with other related approaches.

\section{Thalamocortical Mechanisms for Temporal Integration Learning}

We begin with a summary overview of how the LeabraTI model works, in terms of differential functional roles for superficial and deep layers of the neocortex, and loops through the thalamus, and the temporal dynamics of information flow through this circuit.  Then, we explore each of these elements in greater depth, in relation to available biological and cognitive data.

\subsection{Overview of LeabraTI Model}

\begin{figure}
  \centering\includegraphics[width=6in]{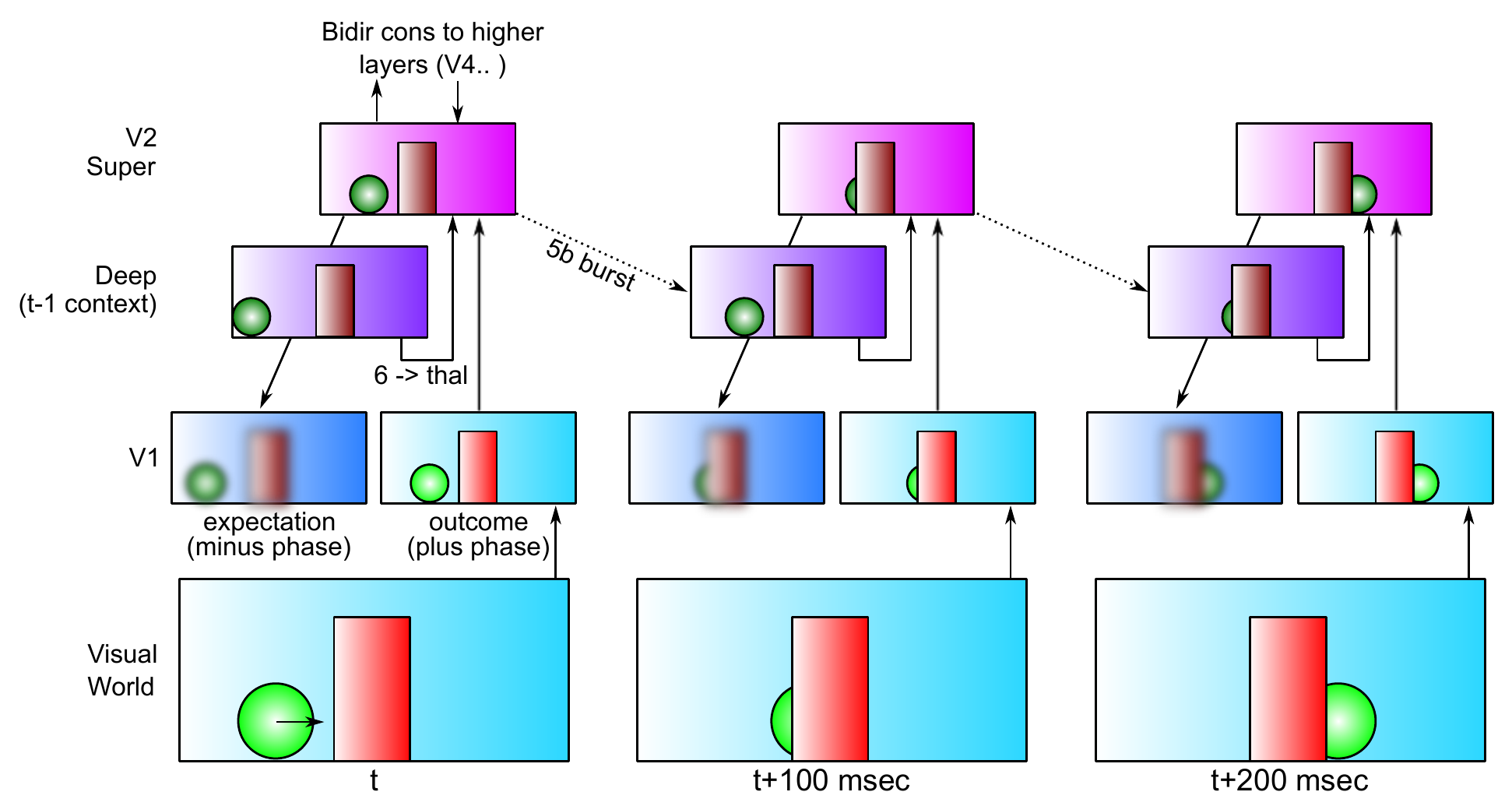}
  \caption{\footnotesize The temporal evolution of information flow in a LeabraTI model predicting visual sequences, over a period of three alpha cycles of 100 msec each.  The  Deep context maintains the prior 100 msec information while the Superficial generates a prediction (in the minus phase) about what will happen next.  Learning occurs in comparing this prediction with the plus phase outcome, which generates an updated activity pattern in the Super layers.  Thus, prediction error is a temporally extended quantity, not coded explicitly in individual neurons.}
  \label{fig.leabra_ti}
\end{figure}

\begin{figure}
  \centering\includegraphics[width=4in]{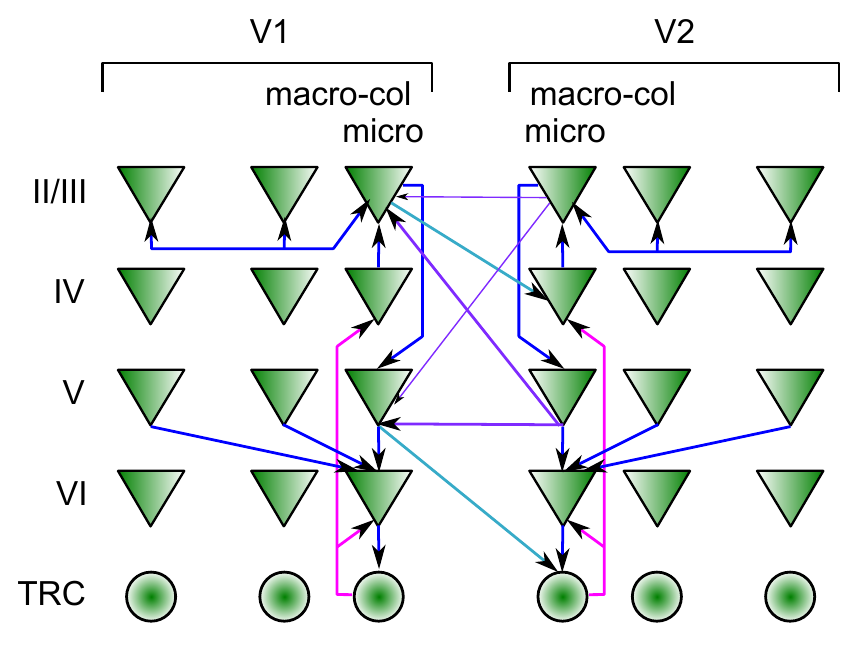}
  \caption{\small Anatomical connectivity supporting the LeabraTI model. Super (II/III) layers have extensive connectivity within and between areas, and do the primary information processing.  Deep layer V integrates contextual information within and between areas, and 5b bursting neurons only update the sustained context, in layer VI, every 100 msec.  These layer VI tonically firing neurons sustain the context through recurrent projections through the thalamic relay cells (TRC), which also communicate the context up to the Super neurons (via IV) to support generation of the next prediction.}
  \label{fig.leabra_ti_bio}
\end{figure}

The LeabraTI model of temporal integration in the neocortex leverages the unique properties of the thalamocortical microcircuit (Figures~\ref{fig.leabra_ti}, \ref{fig.leabra_ti_bio}).  This model makes detailed contact with a wide range of biological and functional data, often with counterintuitive predictions. 
Specifically, the two major claims are:

{\bf 1. Time is discretized into roughly 100 msec intervals}, which correspond to the widely observed alpha rhythm in posterior neocortex.  Computationally, this discretization is important for giving the system sufficient time for bidirectional constraint-satisfaction processing to generate reasonable expectations about what will happen next. Because this processing itself takes time, it is not possible to be continuously generating these predictions, and hence the input must be discretely sampled.  Biologically, the properties of the layer 5b deep neocortical neurons, together with dynamics of thalamic neurons, are thought to underlie the generation of the alpha rhythm \abbrevcite{LorinczKekesiJuhaszEtAl09,FranceschettiGuatteoPanzicaEtAl95,BuffaloFriesLandmanEtAl11,LuczakBarthoHarris13}.  Psychologically, there is increasing evidence for a discretization of perception at the alpha scale \abbrevcite{VanRullenKoch03b}.

{\bf 2. Temporal context is sustained in the deep neocortical layers}, while the superficial layers continuously integrate new information together with this sustained deep context.  Specifically, we argue that the burst-firing dynamics of the layer 5b neurons result in two phases of activity in the deep layers, corresponding to the minus and plus phases of the Leabra algorithm.  The 5b neurons burst fire in the plus phase, driving the updating of downstream layer 6 neurons, which then drive input back down to the thalamus, which then comes back up to the same cortical area, both to layer 6 and up to layer 4 and from there onto the superficial layers (Figure~\ref{fig.leabra_ti_bio}). However, in the minus phase, the 5b neurons are relatively quiescent, and this protects the layer 6 neurons from further influences, enabling them to continue to represent the temporal context from the prior plus phase.  The sustained firing of this layer 6 signal then feeds into the superficial layers, which integrate this prior temporal context with information from all over the rest of the cortex (via inter-areal bidirectional excitatory projections), to produce an expectation about what will happen next.  Then, whatever does happen, happens, and that constitutes the plus phase against which the prior expectation is compared, to drive error-driven learning.  The STDP-based {\em XCAL} learning mechanism in Leabra \cite{OReillyMunakataFrankEtAl12} automatically computes this comparison in a biologically-plausible fashion, in terms of a rapidly adapting threshold between long term depression (LTD) and long term potentiation (LTP).  This minus-plus phase oscillation at the alpha frequency thus constitutes the fundamental ``clock cycle'' of cortical computation, according to this framework.

\subsection{Computational Properties, in Relation to the SRN}

As noted above, at a purely computational level, the use of temporal context in LeabraTI to inform new predictions is very similar to the role of the context layer in a simple recurrent network (SRN) \cite{Elman90,Elman91,Jordan89}.  In effect, we hypothesize that the time step for updating an SRN-like context layer is the 100 msec alpha cycle, and during a single alpha cycle, considerable bidirectional constraint satisfaction neural processing is taking place within a LeabraTI network.  This contrasts with the standard SRN, which is typically implemented in a feedforward backpropagation network, where each time step and context update corresponds to a single feedforward activation pass through the network.  Despite this important difference, and several others that we discuss below, there are some critical computational lessons that we adopt directly from the SRN.

One of the most powerful features of the SRN is that it enables error-driven learning, instead of arbitrary parameter settings, to determine how prior information is integrated with new information.  Thus, SRNs can learn to hold onto some important information for a relatively long interval, while rapidly updating other information that is only relevant for a shorter duration \cite[e.g.,]{CleeremansServanSchreiberMcClelland89,Cleeremans93}.  This same flexibility is present in our LeabraTI model.  Furthermore, because this temporal context information is hypothesized to be present throughout the entire neocortex (in every microcolumn of tissue), the LeabraTI model provides a more pervasive and interconnected form of temporal integration compared to the SRN, which typically just has a single temporal context layer associated with the internal ``hidden'' layer of processing units.

\begin{figure}
  \centering\includegraphics[width=3in]{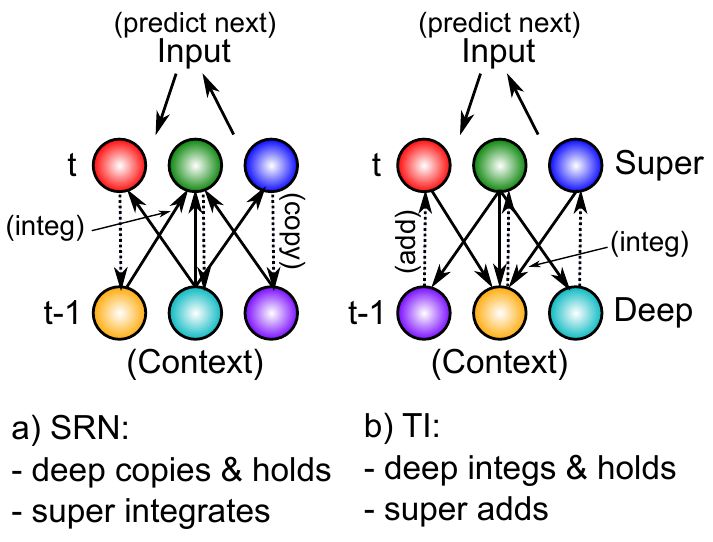}
  \caption{\small How TI computation compares to the SRN mathematically. {\bf a)} In a standard SRN, the context (deep layer biologically) is a copy of the hidden activations from the prior time step, and these are held constant while the hidden layer (superficial) units integrate the context through learned synaptic weights.  {\bf b)} In LeabraTI, the deep layer performs the weighted integration of the soon-to-be context information from the superficial layer, and then holds this integrated value, and feeds it back as an additive net-input like signal to the superficial layer.  The context net input is pre-computed, instead of having to compute this same value over and over again.  This is more efficient, and more compatible with the diffuse interconnections among the deep layer neurons.  Layer 6 projections to the thalamus and back recirculate this pre-computed net input value into the superficial layers (via layer 4), and back into itself to support maintenance of the held value.}
  \label{fig.srn_vs_ti}
\end{figure}

An extensive computational analysis of what makes the SRN work as well as it does, and explorations of a range of possible alternative frameworks, has led us to an important general principle: {\em future outcomes determine what is relevant from the past}.  At some level, this may seem obvious, but it has significant implications for predictive learning mechanisms based on temporal context.  It means that the information encoded in a temporal context representation cannot be learned at the time when that information is presently active.  Instead, the relevant contextual information is learned on the basis of what happens next.  This explains the peculiar power of the otherwise strange property of the SRN: the temporal context information is preserved as a {\em direct copy} of the state of the hidden layer units on the previous time step (Figure~\ref{fig.srn_vs_ti}), and then learned synaptic weights integrate that copied context information into the next hidden state (which is then copied to the context again, and so on).  This enables the error-driven learning taking place in the {\em current} time step to determine how context information from the {\em previous} time step is integrated.  And the simple direct copy operation eschews any attempt to shape this temporal context itself, instead relying on the learning pressure that shapes the hidden layer representations to also shape the context representations.  In other words, this copy operation is essential, because there is no other viable source of learning signals to shape the nature of the context representation itself (because these learning signals require future outcomes, which are by definition only available later).

The direct copy operation of the SRN is however seemingly problematic from a biological perspective: how could neurons copy activations from another set of neurons at some discrete point in time, and then hold onto those copied values for a duration of 100 msec, which is a reasonably long period of time in neural terms (e.g., a rapidly firing cortical neuron fires at around 100 Hz, meaning that it will fire 10 times within that context frame).  Surprisingly, as summarized above and detailed below, the appropriate mechanisms seem to exist: the phasic bursting of the layer 5b pyramidal neurons acts like a kind of ``gate'' for the updating of the temporal context information, which is maintained through the layer 6 thalamocortical recurrent loops, and communicated up to the superficial layers through layer 4.  However, this biology is more compatible with a particular rotation of the SRN-style computation (Figure~\ref{fig.srn_vs_ti}), where the context is immediately sent through the adaptive synaptic weights that integrate this information, and what is fed back through the layer 6 thalamocortical loops is actually the pre-computed net input from the context onto a given hidden unit, not the raw context information itself.  Computationally, and metabolically, this is a much more efficient mechanism, because the context is, by definition, unchanging over the 100 msec alpha cycle, and thus it makes more sense to pre-compute the synaptic integration, rather than repeatedly re-computing this same synaptic integration over and over again.  Specifically, the extensive collateral connectivity among deep layer 5 neurons is what computes this synaptic integration of the context signal, as we review in the next section.

Mathematically, the LeabraTI context computation is identical to the SRN, despite the rotation of how the computation is performed, as described in detail in Appendix A.  However, this rotation does raise other questions about how the learning of the synaptic weights into the deep layers takes place, which we discuss later in the section on Outstanding Questions.  Also, the LeabraTI model differs from the classic SRN in a number of important ways, many by virtue of it being a biologically realistic Leabra model, with extensive bidirectional connectivity, inhibitory competition within layers, and an ion conductance-based activation function \cite{OReillyMunakata00,OReillyMunakataFrankEtAl12}.  We elaborate on these network-level issues next.

\subsection{Network Architecture and Implicit versus Explicit Prediction}

One particularly important difference between LeabraTI and a standard SRN arises because of the bidirectional connectivity in LeabraTI and the temporally distributed error-driven learning mechanism, which allows us to use a single input layer to represent both the expectation and outcome, interleaved across time (as shown in Figure~\ref{fig.leabra_ti}).  Furthermore, higher (deeper) layers in the network contribute both to {\em recognition} of the current inputs, and to the {\em prediction} of the next ones, by virtue of their bidirectional interconnectivity with the lower layers.  In contrast, a standard SRN has a fixed input layer feeding through the hidden layer to generate a prediction of the input at the next time step, over an entirely separate set of output units, with no such bidirectional interactions.  Thus, the LeabraTI model provides a more natural mechanism for predictive learning in the brain: the predictions are generated over the very same layers that represent the outcomes against which those predictions are compared.

However, this issue of interleaving predictions vs. outcomes over an input layer raises some important issues, both computationally and biologically, and we have found that a more {\em implicit} form of prediction (which one could perhaps term {\em anticipation} or {\em preparation}) resolves these issues, while retaining much of the same computational power of the full {\em explicit} prediction framework.  The distinction between explicit and implicit prediction centers around whether the input layer is driven by external input during the minus phase.  For explicit prediction, the input layer in the minus phase can only be driven by top-down activation from within the network itself, which thus must be fully responsible for predicting what will happen in the plus phase, when the external input drives the input layer.  For implicit prediction, we allow the input to be driven externally in the minus phase (as well as the plus phase), but with a dynamic parameterization that makes this input have a weaker impact on the rest of the network in the minus phase compared to the plus phase.  In effect, the implicit form is a graded generalization of the explicit case: the input is only partially predicted in the minus phase.  However, from a computational learning perspective, the network still benefits from the error-driven learning signal based on the difference between these two phases: it will still learn to drive the network state in the minus phase to be as similar as possible to that in the plus phase, which is qualitatively similar to the error signal from the explicit prediction case, just weaker in magnitude.

The main computational problem with full explicit prediction case is that for complex real-world input signals (e.g., visual images impinging on primary visual cortex, as we simulate in  the model described later), we do not think that the network can or should be capable of accurately predicting the next input.  Instead, there are many reasons to believe that a primary function of cortical processing is to actively discard massive amounts of information that comes in through the senses, so that only the most relevant and refined signals are retained and processed further \cite{OReillyMunakata00,OReillyMunakataFrankEtAl12}.  In this case, the problem is that there will be a strong and persistent error signal generated by these fuzzy, incomplete predictions, and that will wreak havoc with the error-driven learning mechanism: these error signals will swamp everything else, constantly driving synaptic weights to extremes, and preventing more subtle information from being learned.  In contrast, the implicit prediction case relieves the network from the burden of predicting every last detail of the input, and merely requires that the internal network state learn to be {\em compatible} with the new inputs.

From a biological perspective, this implicit prediction framework provides a much better fit to what we know about the behavior of neurons in area V1 (primary visual cortex).  Specifically, there is no evidence that the thalamic inputs to V1 (from the LGN) are turned off during the putative minus phase of the alpha cycle.  However, there is a solid basis for a dynamic modulation of the strength of bottom-up signals from V1 to V2 and other higher layers, which comes from the same layer 5b bursting neurons that we think drive updating of the temporal context representations.  These 5b neurons also drive trans-thalamic projection pathways between areas (e.g., V1 to V2), in parallel to the direct cortico-cortical projections from superficial pyramidals.  Thus, receiving neurons in area V2 will experience a continuous input from V1 superficial neurons throughout the minus and plus phases of the alpha cycle, along with an extra phasic burst in the plus phase.  This constitutes the dynamic parameterization mentioned above, and it ensures that more bottom-up signal is present during the plus phase.  The error-driven learning will then drive the extant top-down and recurrent activation during the minus phase to more closely approximate this bottom-up-heavy plus phase signal.  In practice, we capture this differential parameterization with two different multipliers on synaptic connectivity strengths in the minus and plus phases, and we show later that it is highly effective in driving learning in the network.

Next, we address the biological connections of our framework in greater depth, starting with the basic functional neuroanatomy, followed by electrophysiological and behavioral data.

\subsection{Neurobiology of the Thalamocortical Loops}

We start by reviewing some of the key findings regarding the anatomical and physiological properties of the thalamocortical circuits, as summarized in Figure~\ref{fig.leabra_ti_bio}.  We draw heavily upon integrative reviews \cite[e.g.,]{Thomson10,ThomsonLamy07,SchubertKotterStaiger07,ShermanGuillery06,DouglasMartin04,RocklandPandya79}, which establish the following well-accepted conclusions:
\begin{itemize}
\item Activation generally flows in from the thalamus up to cortical layer 4, and from there up to superficial layers 2/3, and then down to deep layer 5, and finally to layer 6, which predominantly receives from layer 5.  Layer 6 in turn projects back down to the thalamus, which reciprocates with projections back up to layer 6, and up to layer 4.  Thus, despite various possible shortcuts along the way, there is a predominant directionality to activation propagation through the circuit. 
\item There are two sublamina of layer 5 neurons, 5a and 5b, and each such sublamina contains regular spiking (RS) and intrinsic bursting (IB) subtypes.  The 5a neurons are distinctive in having extensive intermixing within their own subtype (i.e., 5a projecting to 5a, both within and across columns; \nopcite{SchubertKotterStaiger07}).  In layer 5b, the IB subtype has extensive lateral connectivity, and broad receptive fields, whereas the RS subtype has more focal within-column connectivity \cite{SchubertKotterStaiger07}.  The 5b IB intrinsic bursting dynamics occur typically at around the alpha frequency \cite{FranceschettiGuatteoPanzicaEtAl95,FlintConnors96,SilvaAmitaiConnors91,ConnorsGutnickPrince82}.  Our interpretation of this data is that the 5a neurons help to integrate the contextual net input information (per Figure~\ref{fig.srn_vs_ti}), with the 5b intrinsic bursting neurons also doing considerable integration, and then providing the critical timing for when context is updated (at the end of the plus phase).  And 5b neurons instigate a trans-thalamic projection to higher areas \cite{ShermanGuillery06}, consistent with the implicit prediction framework.
\item Consistent with the hypothesized integrative role, the layer 5 neurons generally exhibit broader, integrative tuning compared to superficial layer neurons \cite{SchubertKotterStaiger07}.
\item There are at least three subtypes of pyramidal neurons in layer 6: two types of corticothalamic (CT) and one type of corticocortical (CC) \cite{Thomson10}.  The upper-layer 6 subtype of CT projects to the thalamic area that reciprocally innervates this cortical area (e.g., LGN for V1), and it exhibits a regular spiking profile.  Our hypothesis is that these regular spiking thalamic projecting neurons continuously rebroadcast the integrated context signal that they just received from their strong 5b input projections (until the 5b neurons burst again in the next alpha cycle).  Interestingly, these CT neurons exhibit facilitating short-term dynamics, unlike all other pyramidal neurons, which exhibit depressing dynamics \cite{Thomson10} --- this is suggestive of a specialized function for this cell type, which fits well with this need for short-term maintenance over the alpha cycle.  The other subtype of CT neurons receive more strongly from layer 6 collaterals \cite{ZarrinparCallaway06}, and project to other thalamic targets (e.g., to secondary areas).  These layer 6 collaterals originate largely from the CC subtype, which receives primarily from deep layers (5a, 5b, 6), and projects almost exclusively laterally to other layer 6 neurons.  Interestingly, these layer 6 CC neurons exhibit a strong intrinsic bursting, rapidly accommodating activity pattern, in contrast to the CT regular spiking --- they might represent an additional contributor to the alpha-phase gating dynamic attributed to 5b neurons, perhaps with different overall connectivity.  In addition to our hypothesized primary function for the CT neurons, recent data suggests they may also be involved in feedback gain normalization \cite{OlsenBortoneAdesnikEtAl12}.
\item Thalamic relay neurons that receive from layer 6 project back up to layer 6, and also up to layer 4, in a focal, reciprocal, point-to-point fashion \cite{ShermanGuillery06,Thomson10}.  We hypothesize that this projection sends the integrated contextual signal for a single microcolumn back up to that same microcolumn, with projections into layer 6 serving to sustain the contextual signal over the ensuing alpha cycle, and those into layer 4 playing the key role of providing the context net input to the integrative superficial layer ``hidden'' neurons.  In addition, layer 6 neurons also project directly to layer 4, and while these projections are relatively weak \cite{HirschMartinez06b}, they do activate a metabotropic glutamate receptor (mGluR) that produces sustained depolarization \cite{LeeSherman09} --- this is another possible route for sustained context information to drive layer 4 firing.
\end{itemize}

In summary, it seems that the somewhat peculiar computational demands of the LeabraTI model fit well with the known features of the thalmocortical circuit.  Furthermore, this framework provides a potential answer to the important question as to why there is all this considerable complexity and differentiation of function within the neocortical layers.  It is likely that at least a few other distinct functional mechanisms are embedded within this circuitry, so there is always more work to be done, but at least the present proposal provides some degree of synthesis of a range of otherwise disconnected biological properties, and a number of further testable predictions that are enumerated in the Discussion section.

\subsection{Electrophysiology of Superficial and Deep Layer Neurons}

\begin{figure}
  \centering\includegraphics[width=6in]{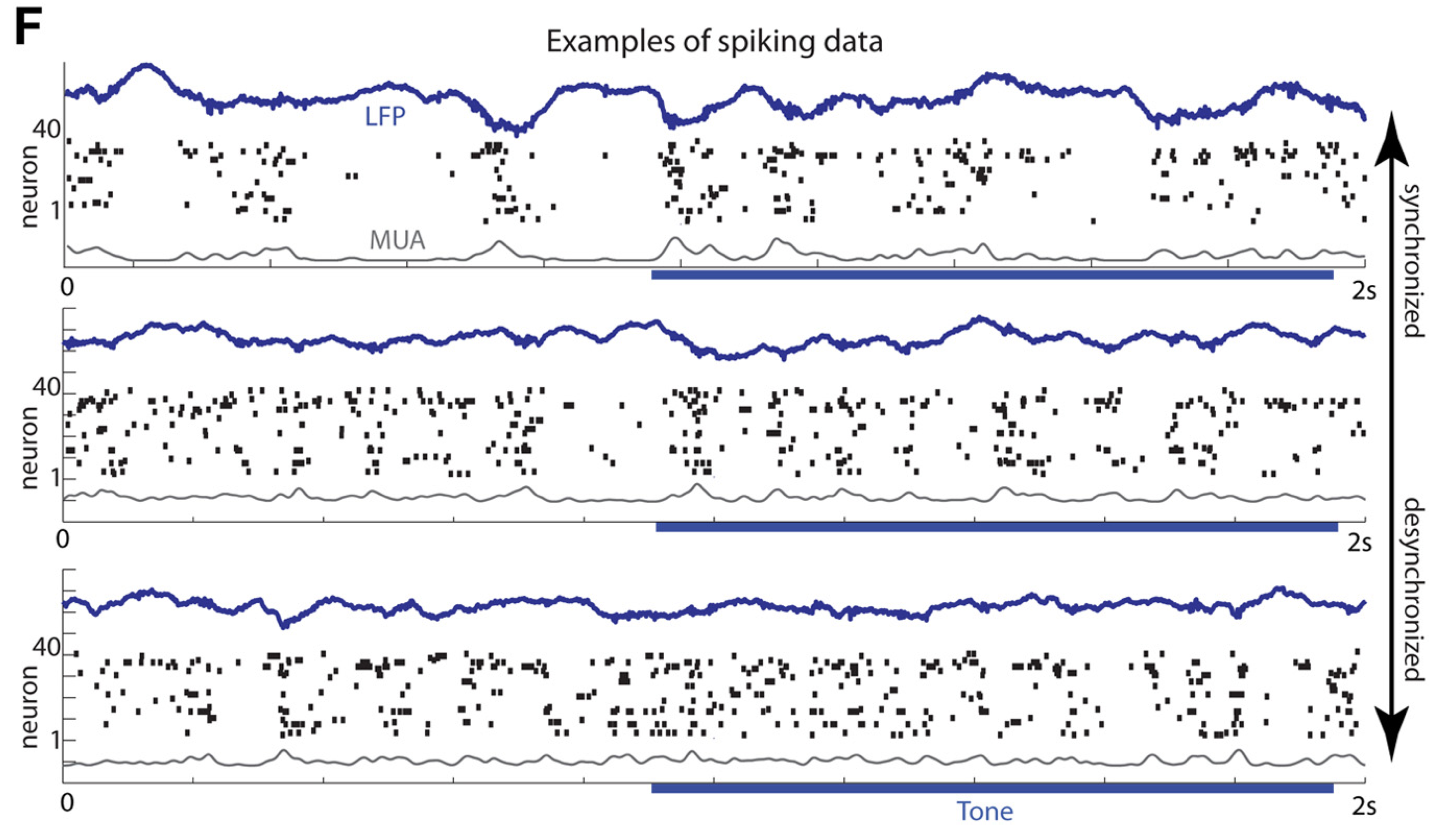}
  \caption{\small Discretization of a continuous tone stimulus in deep layer 5 neurons at the alpha frequency (Figure 6F from Luczak et al, 2013).  This is consistent with the discretized alpha-frequency updating of temporal context representations in the deep layers in the LeabraTI model.}
  \label{fig.sound_disc_alpha}
\end{figure}

Building upon the core functional anatomy reviewed above, we now examine studies that have recorded activity of superficial and deep layer cortical neurons in behavioral tasks, which provides a broader perspective on the functional dynamics of the thalamocortical loops.  For example, one landmark study showed that superficial cortical layers exhibit mainly gamma frequency power (peak $\sim$50 Hz), whereas deep cortical layers exhibit mainly alpha frequency power (peak $\sim$10 Hz) \cite{BuffaloFriesLandmanEtAl11}.  This is consistent with our framework, where the deep layer context representations are updated at the alpha frequency.  More direct evidence comes from a recording of deep layer 5 neurons in auditory cortex during presentation of sustained tone stimuli \cite{LuczakBarthoHarris13}.  These neurons exhibited alpha-frequency segmentation of this continuous input stimulus, just as we would expect from our model (Figure~\ref{fig.sound_disc_alpha}).  Other studies employing depth electrodes to simultaneously record from multiple layers within a patch of cortex have found large differences in the spectral coherence of superficial and deep neurons \cite{MaierAdamsAuraEtAl10}, as would be predicted by LeabraTI.

A similar experimental paradigm expands on these findings by demonstrating cross-frequency coupling between gamma and alpha spectra localized to superficial and deep layers, respectively \cite{SpaakBonnefondMaierEtAl12}. The cross-frequency coupling was characterized by a clear nesting of gamma activity within alpha cycles, suggesting that deep neurons' alpha coherence might subserve a task-independent $\sim$10 Hz duty cycle for continuous integration by superficial neurons.  This suggests that alpha activity (from deep cortical layers) provides the broader temporal context within which superficial layer processing operates \cite{VanRullenKoch03b,JensenBonnefondVanRullen12}.  Finally, there is evidence that the strong 10 Hz coherence of deep neurons persists even with constant sensory stimulation \textit{in vivo}: \incite{MaierAuraLeopold11} found Layer 5 potentials that were not phase-locked to visual stimulation with a strong 10 Hz component as long as the stimulus was present. 

\subsection{Behavioral Evidence of Discretized Perception at the Alpha Frequency}

Next, we turn to behavioral data indicating at least some degree of discretization of perception at the alpha frequency.  The question of whether perception is discrete or continuous has occupied the literature for over 30 years \cite{VarelaToroJohnEtAl81,VanRullenKoch03b,JensenBonnefondVanRullen12}. Our everyday experience suggests that perception is undeniably continuous, but a number of phenomena support the idea that it is discretized at the alpha frequency, at least to some extent.  Critically, the LeabraTI predicts a mixture of both continuous and discrete aspects to perception, because the superficial layers in the model are continuously updating, whereas it is only the deep layers that are discretized at the alpha frequency.  Overall, we find that this mixture of both continuous and discretized aspects provides a better fit to the data than either extreme case alone.

\incite{VarelaToroJohnEtAl81} provided the first demonstration of discretization at the alpha rhythm by presenting two light stimuli with a short, but constant inter-stimulus interval such that they would be perceived as either illuminating simultaneously or sequentially with equal probability. When the illumination was triggered at the peak of an alpha cycle, subjects generally perceived the lights as simultaneous compared to sequentially when they were presented at the trough.  Although subsequent replication of \posscite{VarelaToroJohnEtAl81} results has failed \cite{VanRullenKoch03b}, there are a number of other phenomena that are consistent with discretized perception at the alpha frequency. 

For example, the {\em wagon wheel illusion}, where a rotating spoked wheel appears to switch direction at certain speeds due typically to a strobing light or as a result of discrete movie frames, also occurs under continuous illumination, and is maximal at 10 Hz \cite{VanRullenReddyKoch05,VanRullenReddyKoch06}.  This suggests that the alpha rhythm might impose a similar temporal aliasing effect, due to a frame-like discretization process.  A recent investigation indicated that a static wagon wheel-like stimulus also flickers at rates estimated at $\sim$10 Hz when viewed in the visual periphery outside the scope of overt attention \cite{SokoliukVanRullen13}.  Relatedly, illusory jitter of high-contrast edges has been shown to occur at 10 Hz \cite{AmanoArnoldTakedaEtAl08}. All of these effects have been correlated with increased alpha-band power over the visual cortices and can be accounted for by a model that posits periodic fluctuations in sensory efficacy at approximately 10 Hz.

\begin{figure}
  \centering
  \includegraphics[width=2.75in]{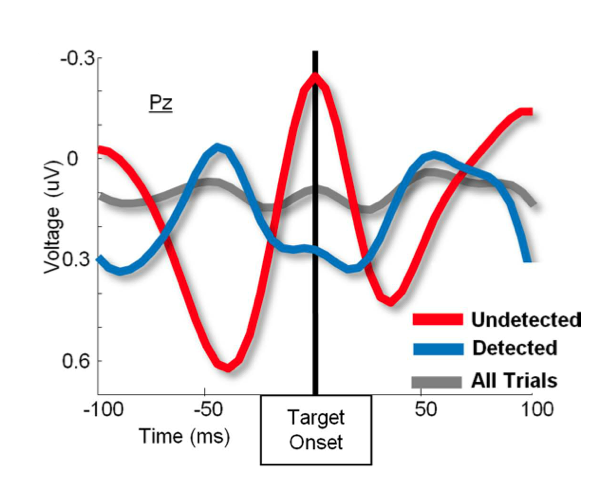}
  \caption{\small Correlation between at-threshold stimulus detection and alpha frequency cycle phase, showing a clear phasic dependency.  Reproduced from \protect\incite{MathewsonFabianiGrattonEtAl10}}
  \label{fig.alpha_phase_onset}
\end{figure}

The phase of ongoing alpha has activity has also been related to sensory processing efficacy. Errors made processing at-threshold stimuli have been suggested to arise from alpha phase at stimulus onset \cite{BuschDuboisVanRullen09,MathewsonFabianiGrattonEtAl10,VanrullenDubois11}. Specifically, analyses that split data based on whether stimuli were successfully perceived have indicated opposing phases for successful versus unsuccessful detection (Figure~\ref{fig.alpha_phase_onset}). For undetected stimuli, the alpha cycle was approximately in phase at target onset and 180$^{\circ}$ out of phase 50 ms later when processing begins in primary visual cortex \cite{NowakBullier97}. LeabraTI suggests that this impairment is specifically due to the prediction computation during the trough (minus phase) of the alpha cycle, when bottom-up inputs are at their weakest.  In contrast, stimuli arriving in time for the plus-phase peak of the alpha cycle obtain a facilitated boost in processing due to the effects of layer 5b bursting on updating context representations and driving transthalamic circuits.

Given the importance of alpha phase in shaping the envelope of successful perception, it seems that there should be a mechanism in place to synchronize the alpha phase to the timing of environmental stimuli.  Indeed, intrinsic oscillations have been shown to phase-lock to exogenous rhythmic visual and auditory stimulation \cite{FujiokaTrainorLargeEtAl09,SpaakdeLangeJensen14,CalderoneLakatosButlerEtAl14}. This phase-locking ensures that environmental events coincide with key neural events that affect sensory efficacy, like layer 5b bursts that update temporal context. Furthermore, higher frequency bands such as gamma are entrained to alpha phase, which causes momentary modulations in their power \cite{LakatosKarmosMehtaEtAl08,SpaakBonnefondMaierEtAl12}. \incite{MathewsonPrudhommeFabianiEtAl12} presented subjects with a train of stimuli that were either rhythmic, and thus reliably predicted the temporal onset of a masked probe, or were arhythmic and unpredictable. Rhythmic stimulus trains caused entrained endogenous alpha oscillations and as a result, probes that occurred in either 100 ms or 200 ms after the probe were less susceptible to the effects of masking, due to their processing falling within the peak of the next alpha cycle. fMRI studies have indicated that hemodynamic responses are sensitive to alpha phase \cite{ScheeringaMazaheriBojakEtAl11}. For example, presenting facial motion at alpha-band frequencies results in higher overall responses in face-selective regions \cite{SchultzBrockhausBulthoffEtAl13}. Alpha phase might also play a role in gating long-range functional connectivity between visual processing areas \cite{HanslmayrVolbergWimberEtAl13}. 

Phase resetting is thought to underly the alpha rhythm's environmental phase-locking properties \cite{CalderoneLakatosButlerEtAl14}. Phase resetting also provides flexibility for the rhythm to align with unexpected salient stimuli that capture attention. For example, salient flashes can cause fluctuations in perceptual efficacy that oscillate at 10 Hz after the flash onset \cite{LandauFries12}. Salient sounds can also cross-modally reset alpha in the visual cortices with a similar fluctuations in accuracy after the event \cite{FiebelkornFoxeButlerEtAl11,RomeiGrossThut12}. In an effect reminiscent of the original simultaneity/sequentiality paradigm of \incite{VarelaToroJohnEtAl81}, \incite{ShamsKamitaniShimojo02} showed that salient sounds played during a persistent stimulus can cause the perception of multiple flashes of the stimulus. The framework developed here suggests that this illusory perception is due to the cross-modal reset of the alpha rhythm by sound, which begins with a trough during which visual information from the sensory periphery is suppressed in favor of prediction, creating an illusory gap in the persistent visual stimulus. 

There are a handful of other phenomena that might be related to discretization of perception specifically at alpha frequency. The heavily-studied attentional blink, for example, occurs when observers respond to the onset of a target in a stream of stimuli presented at a constant rate (typically around 10 Hz). Successful detection of the target causes impairment of subsequent targets for several hundred milliseconds. These results can at least partially be accounted for by alpha entrainment and power/phase properties after the onset of the first target, and recent data using EEG recording to quantify these properties support this idea \cite{JansonDeVosThorneEtAl14,ZaunerFellingerGrossEtAl12}. Eye movements also elicit fluctuations in sensory efficacy that bear a highly similar profile to those elicited by the alpha rhythm. Saccades are characterized by extreme suppression of visual activity and visual processing following a saccade is enhanced relative to baseline \cite{MelloniSchwiedrzikRodriguezEtAl09,ParadisoMeshiPisarcikEtAl12}. Monkey electrophysiology has indicated that this is due to a synchronization of 
neural activity 100 ms after a fixation \cite{MaldonadoBabulSingerEtAl08}, likely from a phase reset in one or more frequency bands \cite{RajkaiLakatosChenEtAl08,ItoMaldonadoSingerEtAl11}.

To summarize the data reviewed here, there are a number of visual phenomena that show periodicities in perceptual processing, most of which fall in the 10hz alpha band. Furthermore, mechanisms exist to reset and lock the phase of alpha to important environmental stimuli. In the context of LeabraTI, this phase alignment is necessary for ensuring that important environmental inputs are coded during the peak of the alpha cycle (LeabraTI's plus phase) at which point neuronal excitability is at its highest.

\begin{figure}
  \centering
  \includegraphics[width=3.25in]{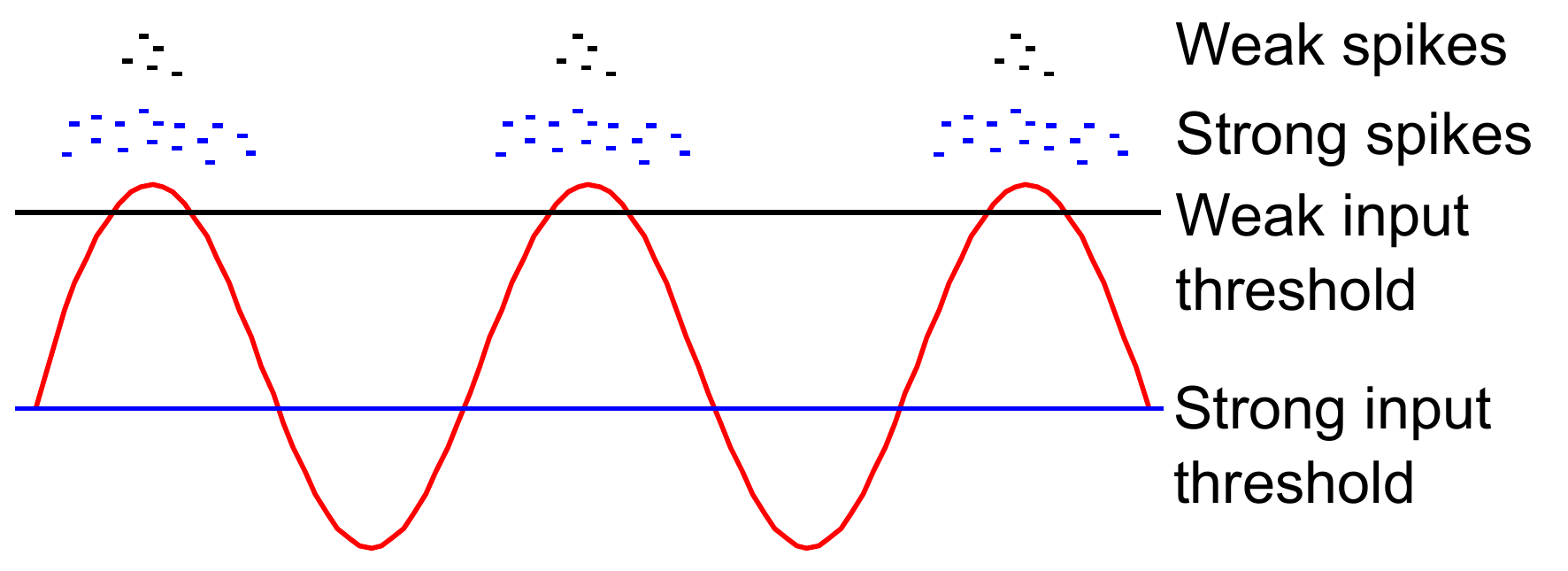}
  \caption{\small Effect of level of synaptic input drive on measured power of a given oscillatory frequency: stronger synaptic input results in a lower effective firing threshold compared to the oscillation dynamics (represented by the red sine wave), and thus drives neural firing at a wider range of phases in the cycle, reducing measured power.  Only if the synaptic input is strongly phase locked can this dynamic be avoided.}
  \label{fig.oscillation_str_phase}
\end{figure}

One remaining question concerns the relationship between various cognitive states and alpha power. For example, alpha power tends to decrease with increases in attention \cite{MathewsonLlerasBeckEtAl11,JensenBonnefondVanRullen12}.  This is consistent with the general finding that lower-frequency oscillations are enhanced in inverse proportion to information processing engagement (e.g., these oscillations are greatest during sleep and while the eyes are closed, and become less pronounced with attentive processing; \nopcite{NiedermeyerLopesDaSilva05}).  Mechanistically, this occurs because the oscillations are entrained in part by intrinsic membrane properties of neurons, which are more influential when synaptic drive is weaker (Figure~\ref{fig.oscillation_str_phase}).  Put another way, stronger synaptic drive is able to break through the underlying low-frequency modulation for a greater percentage of the cycle, thus reducing the overall measured power of the low-frequency band.

This reduction in alpha with greater processing engagement may seem contradictory to the LeabraTI framework, where you might expect the opposite to hold, given the overall importance of alpha modulation for information processing in this framework.  However, it is essential to carefully consider the full set of circuits, and how they are measured in a given imaging paradigm, to make clear predictions from the LeabraTI framework.  First, as emphasized above, the alpha signal is driven by the deep layer 5b neurons, and the extent to which alpha dominates the overall power spectrum depends on the relative impact of these neurons compared to the superficial neurons, which are thought to be continuously firing.  If they have weak synaptic drive, then these superficial neurons will likely reflect some of the 5b drive.  But as the drive on the superficial neurons increases, they should exhibit less alpha energy, for the basic neurophysiological reasons just described (Figure~\ref{fig.oscillation_str_phase}).  The same arguments hold for all the other neurons in the system (e.g., 5a and 6 regular spiking neurons).  Thus, the basic underlying alpha modulation, which we think persists across all states of cortical processing and arousal levels, will nevertheless be drowned out and less pervasive during states of high arousal and attention.  Even lamina-specific recordings will be contaminated by a mixture of neuron types.  Thus, the true test of the LeabraTI model requires detailed neuron-specific recordings and tracking of how information signals across these neurons update as a function of the overall alpha cycle.

\section{The Computational Benefits of Predictive Learning in Early Vision}

We now turn to a computational exploration of LeabraTI learning that illustrates one important way in which predictive learning can unfold in an ecologically and cognitively plausible fashion, while also imparting clear functional benefits in a well-controlled comparison against a system that does not leverage predictive learning.  Specifically, we build upon our existing work on invariant object recognition \cite{OReillyWyatteHerdEtAl13} by investigating the impact of predictive learning on shaping the early visual representations upon which higher-level object recognition builds.  In our prior model, we found that complex cluttered backgrounds substantially impaired recognition performance compared to objects on blank backgrounds, and this is  true of machine vision algorithms more generally, which hover around 60-70\% accuracy on recognition tasks with complex backgrounds --- this is well below human performance on such tasks.  Here, we investigate the hypothesis that figure-ground segregation processes operating in early visual areas (i.e., areas V2 and V3; \nopcite{QiuVonDerHeydt05}) play a critical role in enabling the high performance of human vision.  A scene is first parsed into a structured encoding of the relevant surfaces, organized in relative depth, and then object recognition operates on relevant subsets of such surfaces, effectively filtering out the irrelevant background features.  In particular, we hypothesize that people (and other mammals) learn these early visual representations by simply predicting how a visual scene will unfold over time, which drives development of representations of the separable surface elements, and how they move over time.

\begin{figure}
  \centering \includegraphics[width=4in]{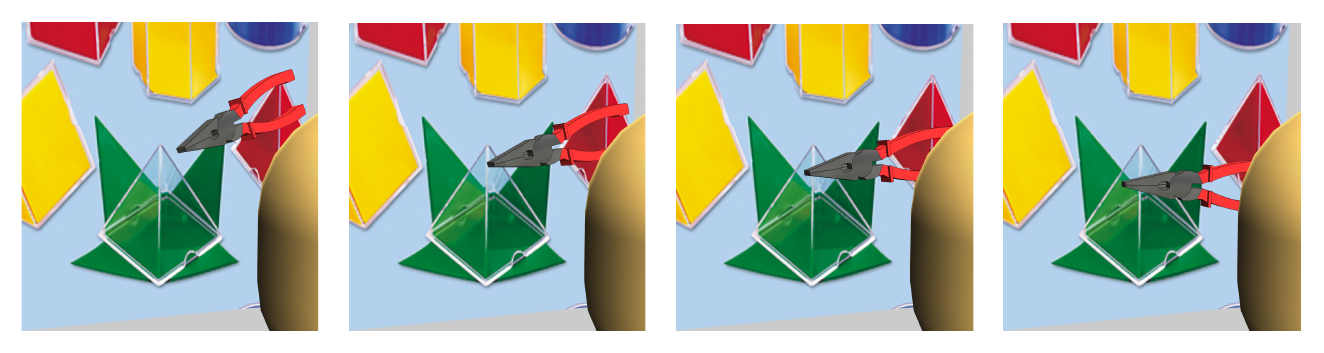}
  \caption{\footnotesize Example of tumbling object (pliers) in front of complex background, used to train LeabraTI model, from binocular eyes of ``emer'' virtual robot.  The model predicts input in the next alpha cycle, learning from prediction errors.}
  \label{fig.tumbling_objs}
\end{figure}

\begin{figure}
  \centering \begin{tabular}{ll}
  A) & B) \\
  \includegraphics[width=2in]{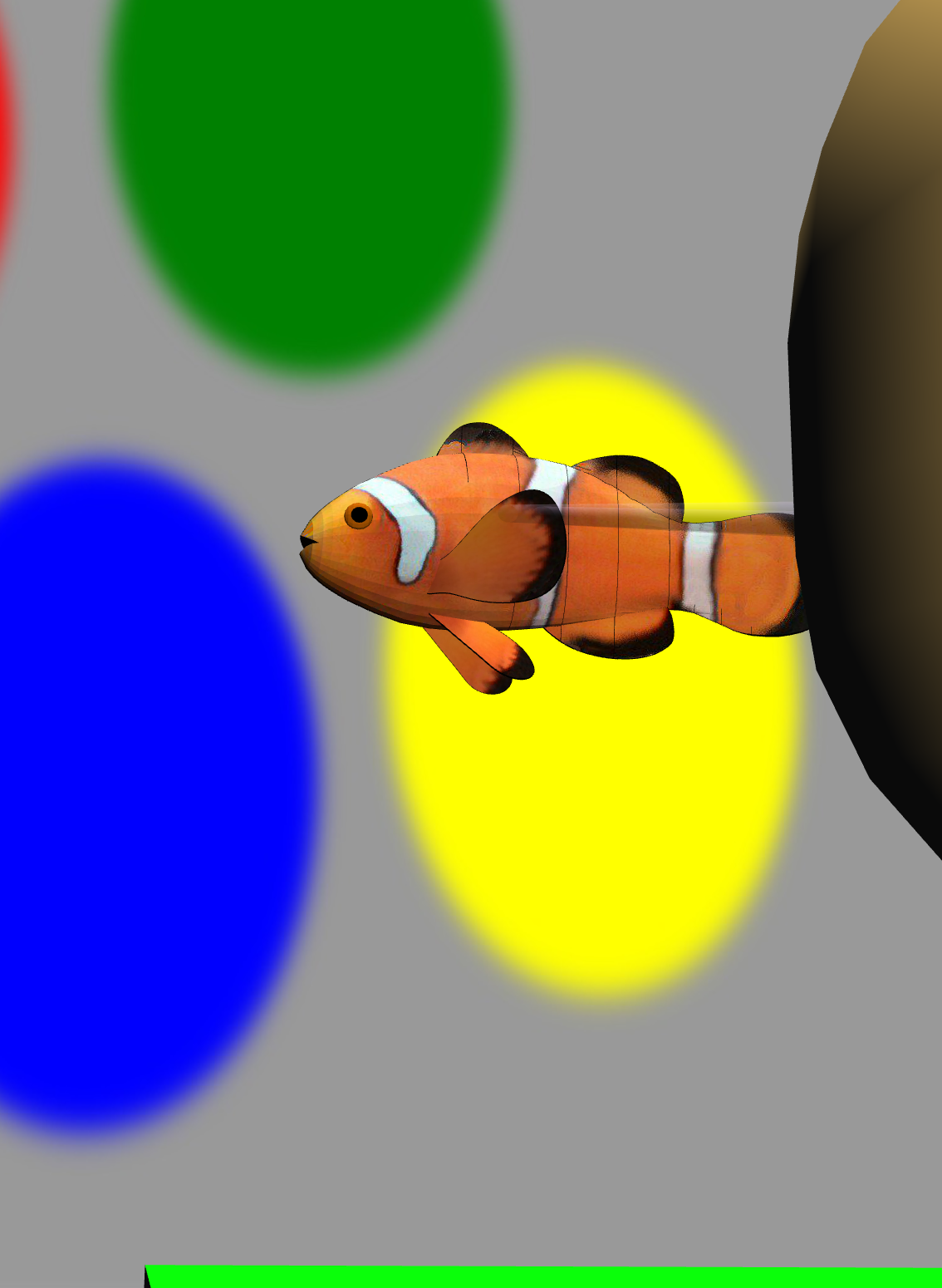} &
  \includegraphics[width=2in]{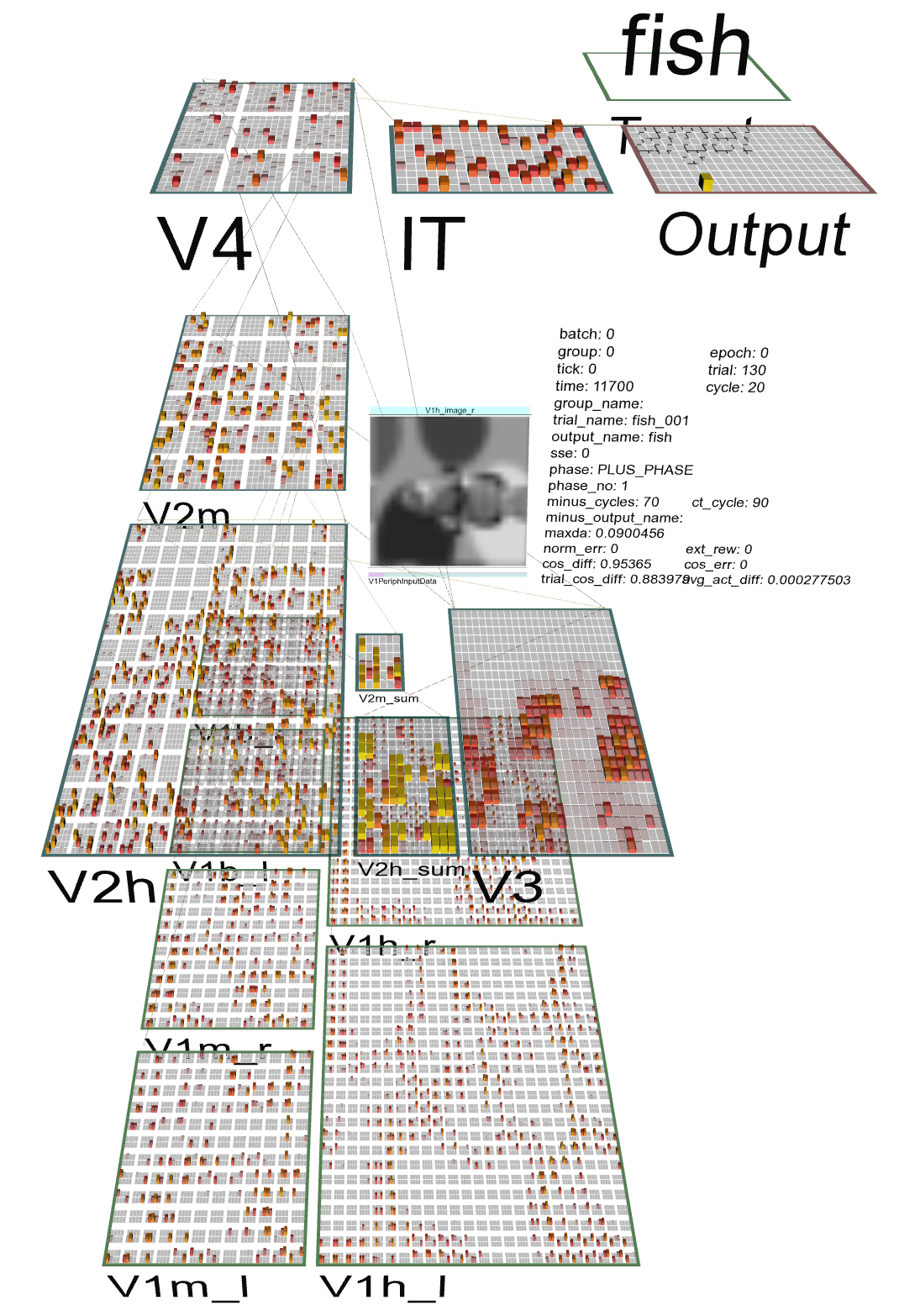}
  \end{tabular}
  \caption{\footnotesize {\bf A)} Virtual environment showing view of virtual agent, looking at a tumbling 3D object (a clown fish in this case), against a complex background.  {\bf B)} The full TI-based LVis network, processing the scene of the fish against background.  Binocular left (l) and right (r) channels across medium (m) and high (h) spatial frequencies are represented across different retinotopically mapped V1 sub-layers, including color-contrast blob layers (V1b).  The challenging nature of the figure-ground problem is evident through visual examination of the activity patterns on V1 -- it is very difficult to determine where the object is compared to the background shapes.  These V1 layers project into corresponding V2 layers, which then project in a bifurcating path to V3 (dorsal pathway) and V4 to IT and verbal Output (ventral object recognition pathway). }
  \label{fig.objrec_ti_net}
\end{figure}

To test whether LeabraTI can learn to encode the structure of visual scenes, we presented our model with rendered 3D movies of 100 different objects from our CU3D 100 object data set (\verb\http://cu3d.colorado.edu\; \nopcite{OReillyWyatteHerdEtAl13}), with each object tumbling through space in front of complex visual backgrounds (Figure~\ref{fig.tumbling_objs}).  These movies were rendered online as the model learns, with randomly-generated motion parameters, backgrounds, objects, etc, so there is a high level of variability and broad, combinatorial sampling of the space.  Replicating the known developmental maturation of areas \cite{ShragerJohnson96}, we start with just the lower visual areas for this initial training (V1, V2, and V3), and then add higher areas to perform object recognition based on these early visual representations.  V1 has two different spatial frequency channels (high: 24x24 locations with 4 oriented edge features x 2 polarities per location; and medium: 12x12 locations with same filters, plus blob-like filters that encode color contrast information without orientation tuning), and V2 samples a 4x4 retinotopically organized window into the corresponding V1 (Figure~\ref{fig.objrec_ti_net}).  V3 is bidirectionally connected to the full extent of V2, to enable it to develop flexible higher-order representations across the whole visual field, and it also has an important topographic projection from a layer that summarizes the overall activity pattern over V2 ({\em V2 sum}), which helps V3 develop topographically organized representations.  We suggest that this V2 sum layer corresponds to the pulvinar of the thalamus, due to it having similar overall properties \cite[e.g.,]{SaalmannPinskWangEtAl12}.  For more details on the structure of the model, see the Appendix.

\begin{figure}
  \centering \includegraphics[width=3in]{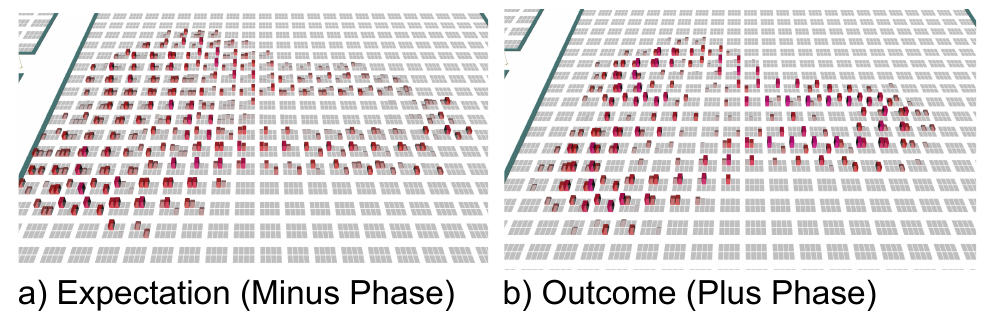}\\
  \caption{\footnotesize Example of expectation vs.\/ outcome for explicit prediction network on higher-res V1 layer for a wrench tumbling through space (with no background) (.56 cosine).}
  \label{fig.objrec_ti_exp_pred}
\end{figure}

We ran the model in both the explicit and implicit prediction modes (as described earlier), each of which provided useful informative results.  In the explicit prediction mode, the network attempted to generate in the minus phase the entire V1 representation for the next time step.  This allows us to compute the overall accuracy of the network as result of training, to determine how much of this challenging task it can actually perform.  Despite the highly variable inputs, we found that the model learns to an overall average cosine of 0.56 to predict the next input frame.  A perfect prediction would be 1.0, so this is well below that, but nevertheless, the cosine is computed over every oriented edge feature in the V1 encoding of the image, so we do not expect anywhere near perfection with that level of detail.  Visually, it is clear that the minus phase predictions capture the overall shape and motion of the objects quite well (Figure~\ref{fig.objrec_ti_exp_pred}).

However, the synaptic weights learned in the explicit prediction mode were highly saturated due to the strong, persistent error signal, and this error signal interfered with the ability of the network to learn the subsequent object recognition task.  These limitations are remedied by the implicit prediction version.  However, with this version, we cannot say exactly how well the network is able to predict the next input.  Nevertheless, we can track the phase-based differences across the V2 and V3 layers, and observe that they decrease systematically over training (indicating that the network is correctly anticipating the plus phase state in the minus phase), while avoiding degenerate representations where this would be trivially true (e.g., having the same pattern active at all times, or no pattern active at all).  Furthermore, when we then add the object recognition task, we find that learned weights do a much better job in filtering out cluttered backgrounds than those trained with explicit prediction learning.

\begin{figure}
  \centering\includegraphics[width=2in]{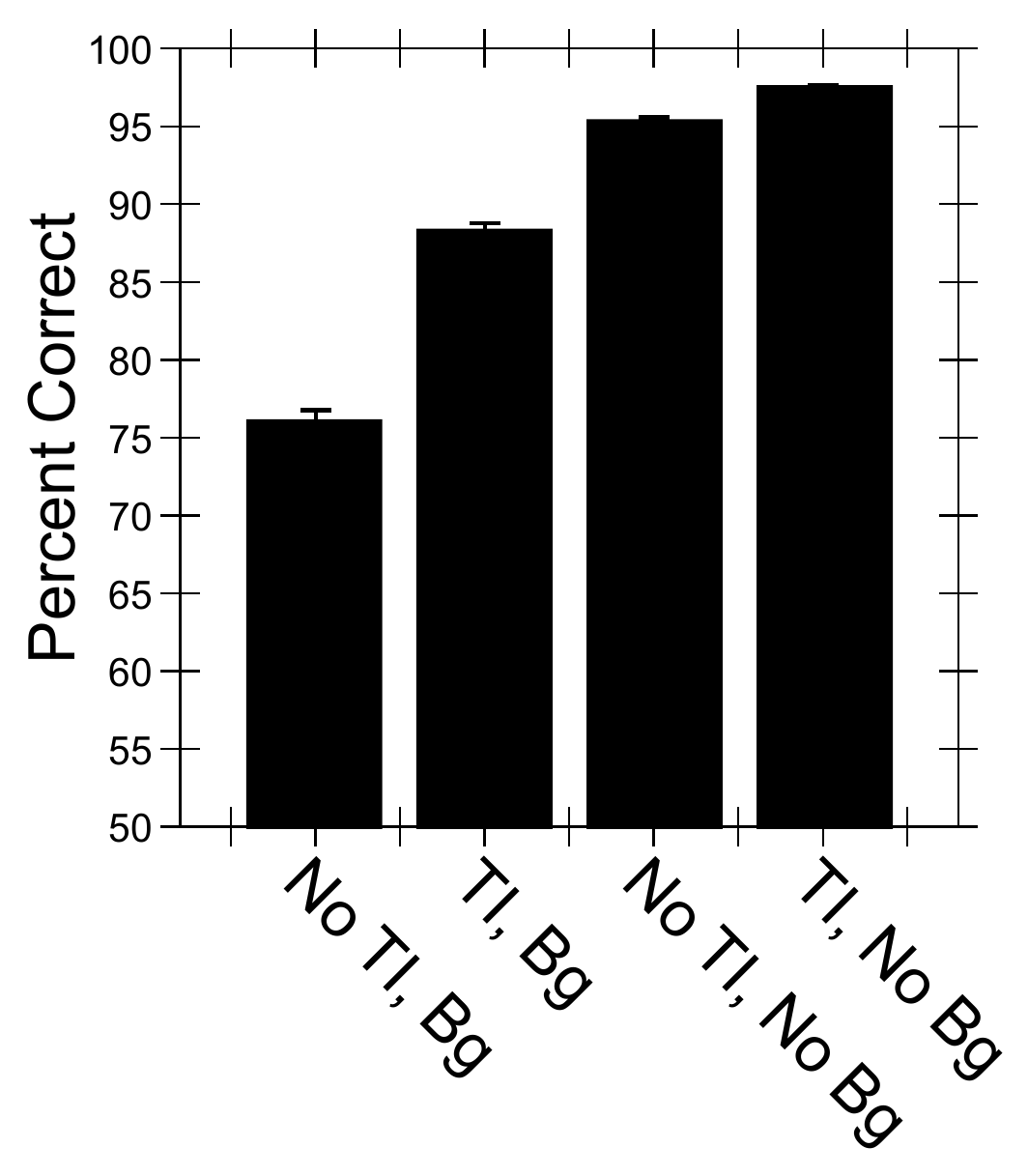}
  \caption{\footnotesize Object recognition accuracy for four training conditions: {\bf No TI, Bg}: no LeabraTI training of V2, V3 on tumbling objects, with backgrounds; {\bf TI, Bg}: with LeabraTI V2, V3 training, and backgrounds; {\bf No TI, No Bg}: no LeabraTI and no backgrounds; {\bf TI, No Bg}: Leabra TI without backgrounds.  TI produces a substantial benefit in recognizing objects with complex backgrounds, and a smaller benefit for the no-background case, consistent with the idea that this TI training is developing good representations of the visual structure of a scene.  Results are from N=5 networks per condition, error bars show SEM.}
  \label{fig.objrec_ti_results}
\end{figure}

After 25,000 trials of visual predictive learning, we then expanded the network to include the higher-level object recognition layers (V4, IT; Figure~\ref{fig.objrec_ti_net}), and trained the network to label the foreground object, as in our previous LVis networks.  In these initial tests, we used only 10 objects, with a reduced range of tumble, to ensure good overall recognition performance from our initial small-scale model.  We also compared this {\em TI} network with others that had no initial predictive learning experience (labeled {\em No TI}), and others that had no backgrounds during object recognition training ({\em No Bg}), to be contrasted against the {\em Bg} case with backgrounds.  The results (Figure~\ref{fig.objrec_ti_results}) show that the initial predictive learning experience produces a substantial improvement in object recognition performance (approaching 15\%), compared to the network with no such experience.  Furthermore, the resulting {\em TI} performance was closer overall to that of the networks without backgrounds, than it was to the {\em No TI} model, indicating that the TI learning had gone a reasonable portion of the way toward extracting the figure objects from the backgrounds.  Finally, the trained {\em TI} weights did also benefit performance for the {\em No Bg} condition model as well, compared to the {\em No TI, No BG} condition.  However, this benefit is much smaller than that on the {\em Bg} condition.

Overall, these results are very encouraging, and suggest that implicit predictive learning in the LeabraTI framework may provide an important approach to learning the 3D and dynamic structure of the visual world, over the early layers of the visual system.  This is an important advance compared to the existing models in this domain, which generally rely on hand-coded representations \cite{Zhaoping05,CraftSchutzeNieburEtAl07}.  We think that the representations that develop through predictive learning will be more powerful and capture a much broader range of structural constraints and regularities in the visual world.  In part, this belief is based on our general failure to find object recognition advantages for our implementations of existing hand-coded figure-ground representation structures (unpublished research), as compared to the significant advantages found here for the TI network.  Despite these encouraging results, considerable work remains to be done -- in particular with respect to the complexity of the scenes that are learned by the TI mechanism, and thus the corresponding amount of visual structure that can be learned.  

\section{Discussion}

In summary, we think that every 100 msec the neocortex is integrating prior context, captured in deep layers across the whole system, to generate a systematic set of implicit predictions across all areas, which is then followed immediately by a wave of alpha-peak activation that reflects the actual perceptual inputs.  This alpha burst wave, which serves as a plus phase for error-driven learning in comparison to the prior prediction (minus phase), is propagated from the periphery inward throughout the rest of the cortex, in a way that depends in part on the layer 5b intrinsic bursting dynamics. The deep-layer context state is updated during this alpha wave, and it then persists through to the next predictive phase, and enables the subsequent predictions to integrate prior context in an effective way. The alternating phases within the alpha cycle drive neocortical neurons to develop more accurate predictions over time, and thereby extract powerful internal representations of the systematic structure of events over time in the environment.

We demonstrated that this predictive learning mechanism with temporal context integration, implemented in the LeabraTI framework, can learn to predict the visual appearance of objects tumbling through space, thereby extracting systematic internal representations of these objects in lower-level visual areas (V2, V3).  These lower-level visual representations provided a solid foundation for subsequent object recognition in higher visual areas (V4, IT), specifically in distinguishing figure from background.  We argue that this is a reasonable (if simplified) model of early visual development, providing an appealing explanation for what babies are doing during the first several months of life, and how this then leads to such rapid learning later.  In future work, we plan to address the extent to which this predictive learning can account for the rapid appearance of basic knowledge about the physical world, which has been argued to be innate \cite[e.g.,]{Spelke94,KellmanSpelke83}.

In the remainder of the discussion, we compare this framework with other related frameworks, and then highlight a few of the many central testable predictions from our model, followed by a discussion of a number of unresolved questions for future research.

\subsection{Comparison with other Frameworks}

\subsubsection{Friston's Free Energy Model}

Perhaps the most closely related model that attempts to make direct contact between a computational learning theory and detailed properties of the neocortex, across a range of different phenomena, is Friston's free energy model \cite{Friston05,Friston10,BastosUsreyAdamsEtAl12}.  This model is based on minimizing free energy, which amounts to minimizing the discrepancy between the predictions of an internal model of the world, and what the world actually presents to the senses.  It leverages the Bayesian generative model framework, and has been offered as a unifying principle for a wide range of phenomena \cite{Friston10}.  At a very broad level, the focus on generative learning is compatible with the predictive learning in LeabraTI, so many of the general arguments in favor of this form of learning are shared between these frameworks.

However, the more detailed claims about mechanism differ significantly between the two.  For example, Friston's model posits the existence of a subset of neocortical neurons that explicitly encode the error between the top-down prediction and the bottom-up input, whereas in LeabraTI this error is only implicit in the temporal difference of activations across the minus vs. plus phase.  Furthermore, their model suggests that feedback projections are predominantly inhibitory, such that the top-down prediction is essentially subtracted from the bottom-up input to compute this error signal.  In support of this model, they highlight data showing that neural responses to unexpected inputs are generally larger than to expected or predicted ones \cite[e.g.,]{SummerfieldTrittschuhMontiEtAl08,BastosUsreyAdamsEtAl12}.  However, this data can also be explained within the LeabraTI framework, where expected inputs result in sharper, more focused neural representations that are consistent across the minus and plus phases, whereas unexpected inputs result in divergent patterns of activity across these phases, that will likely result in a larger number of neurons overall becoming activated (e.g., if there was (implausibly) no overlap at all between the expectation and outcome pattern, there would be roughly two times as many neurons active).  These patterns of neural firing can be discriminated in the electrophysiology, but are difficult to distinguish in fMRI.

More generally, a comprehensive review focused on Bayesian models of the brain concluded that the available evidence from direct neural recordings was not compatible with the error coding idea \cite{KerstenMamassianYuille04}.  Instead, this review concluded that the evidence was more compatible with a role for top-down connections in sharpening and focusing lower-level representations \cite{LeeYangRomeroEtAl02}, which is directly compatible with the excitatory constraint-satisfaction dynamic in LeabraTI.  Furthermore, the idea that feedback projections are inhibitory is at odds with the basic anatomy, where all long-range connections in the neocortex are excitatory.  The excitatory nature of these top-down connections is compatible with the well-supported biased-competition model \cite{DesimoneDuncan95} --- Friston and colleagues attempt to reconcile their inhibitory top-down projections with both the anatomy and these attentional effects, but the explanation requires a number of relatively speculative steps, and the overall system ends up being rather complex overall \cite{BastosUsreyAdamsEtAl12}.

Perhaps the most important difference between these frameworks is that LeabraTI includes a specific mechanism for maintaining and integrating temporal context information (to support predictions based on recent prior history), whereas the free energy model does not address this issue directly.  As emphasized above, this temporal context information is essential for making accurate predictions, and the various significant computational constraints on how these context representations interact with learning suggest that this functionality places significant constraints on the way it must be implemented in the biology.  Thus, we consider it to be a central computational function that any predictive learning framework should address.

\subsubsection{Deep Neural Networks}

As noted in the introduction, there has been a resurgence in error backpropagation and other forms of error-driven learning in deep neural networks (having many hidden layers), driven in part by the impressive gains in performance that these models have demonstrated \cite{CiresanMeierGambardellaEtAl10,CiresanMeierSchmidhuber12,BengioCourvilleVincent13}.  Some of the earlier versions of these models used an incremental learning process for training nested autoencoders, where each additional layer learned to encode the information present in the previous layer \cite{HintonSalakhutdinov06,BengioLamblinPopoviciEtAl07}.  This generative autoencoder framework is generally compatible with the predictive learning in LeabraTI, and we also have found that incrementally training layers (which is compatible with the known developmental progression in cortical plasticity; \nopcite{ShragerJohnson96}) results in better overall performance.  However, more recent models, including those developed by Hinton and colleagues, show that the best performance results from a pure feedforward backpropagation network, combined with a number of important extra ``tricks'' \cite{CiresanMeierGambardellaEtAl10,CiresanMeierSchmidhuber12,KrizhevskySutskeverHinton12,BengioCourvilleVincent13}.  These extra tricks include biologically-supported properties present in Leabra (e.g., winner-take-all learning and sparse representation pressure), and biologically implausible but computationally powerful techniques, most importantly the sharing of synaptic weights across all the neurons in a given layer (known as a {\em convolutional} network).  In general, this work has not been particularly concerned with biological plausibility issues.

Nevertheless, the impressive computational power of these deep neural networks based on error backpropagation provide an important demonstration that the core learning principles built into Leabra can solve challenging problems that we know the neocortex actually solves.  Furthermore, the bidirectional connectivity present in Leabra, and not in these other deep networks, provides an additional computationally powerful mechanism, although it also incurs a significant computational cost, making it more difficult to achieve the large scales used in the purely feedforward models.  We are optimistic that a scaled-up version of the object recognition model presented in this paper will demonstrate the power of predictive learning in LeabraTI, over and above the power of purely error-driven learning in object recognition.  

\subsubsection{Hawkins' Model}

The importance of predictive learning and temporal context are central to the theory advanced by Jeff Hawkins \cite{HawkinsBlakeslee04}.  This theoretical framework has been implemented in various ways, and mapped onto the neocortex \cite{GeorgeHawkins09}.  In one incarnation, the model is similar to the Bayesian belief networks described above, and many of the same issues apply (e.g., this model predicts explicit error coding neurons, among a variety of other response types).  Another more recent incarnation (described apparently only on a white paper available on the website \verb\numenta.org\) diverges from the Bayesian framework, and adopts various heuristic mechanisms for constructing temporal context representations and performing inference and learning.  We think our model provides a computationally more powerful mechanism for learning how to use temporal context information, and learning in general, based on error-driven learning mechanisms.  At the biological level, the two frameworks appear to make a number of distinctive predictions that could be explicitly tested, although enumerating these is beyond the scope of this paper.

\subsubsection{Granger's Model}

Another model which has a detailed mapping onto the thalamocortical circuitry was developed by Granger and colleagues \cite{RodriguezWhitsonGranger04}.  The central idea behind this model is that there are multiple waves of sensory processing, and each is progressively differentiated from the previous ones, producing a temporally-extended sequence of increasingly elaborated categorical encodings ({\em iterative hierarchical clustering}).  The framework also hypothesizes that temporal sequences are encoded via a chaining-based mechanism.  In contrast with the LeabraTI framework, there does not appear to be a predictive learning element to this theory, nor does it address the functional significance of the alpha frequency modulation of these circuits.

\subsection{Predictions}

A paper on the importance of predictive learning certainly must include a section on predictions from this framework!  As in predictive learning, enumerating predictions from a theory provides a way of testing internal representations and refining them in light of observed data.  There are so many possible predictions from our framework, and a good deal of the existing data has already been discussed above, so here we highlight a few important tests that could potentially be made using advanced new optogenetic manipulations and laminar neural recording techniques.

\begin{itemize}
\item If it were possible to selectively block the 5b intrinsic bursting neurons, or perhaps disable their bursting behavior in some other way, we would predict that this would have a significant impact on any task requiring temporal integration of information over time.  For example, discriminating different individuals based on their walking motion, or recognizing a musical tune.  More generally, if any person was brave enough to attempt taking a pharmacological agent that selectively interfered with 5b bursting, we would predict that it would significantly disrupt the basic continuity of consciousness --- everything would feel more fragmented and discontinuous and incoherent.  Indeed, perhaps certain existing psychoactive substances can be understood in part in terms of their modulation of alpha bursting?
\item Neocortical learning should also be significantly impaired with blockage of 5b intrinsic bursting dynamics, because these contribute to the hypothesized plus phase of learning.  To test this prediction, the widely-used statistical learning paradigm would be ideal, where sequences of tones or visual stimuli are presented, with various forms of statistical regularities \cite[e.g.,]{AslinSaffranNewport98}.
\item Using large-scale lamina-specific neural recording techniques, it should be possible to quantify the information encoded in the layer 6 regular spiking (RS) neurons during the trough of the alpha cycle, compared to the information in the superficial layers during the previous alpha peak.  Because we think that the layer 6 RS neurons convey the temporal context information from the prior alpha peak, these two should be more strongly correlated in their information content, as compared to for example the information in superficial layers during the subsequent alpha peak.
\end{itemize}

\subsection{Unresolved Issues and Future Research}

There are a number of important unresolved issues that could be addressed in future research, which are enumerated here.
\begin{itemize}
\item Computationally, the projections into the deep layers should learn according to the plus -- minus phase error signal experienced over the superficial layers.  Furthermore, the same is true in principle for projections into layer 4 neurons.  In our LeabraTI model, we treat all of these pathways as part of a unified microcolumn-level unit, where for example the rate code activation of an individual simulated LeabraTI unit is thought to correspond to the instantaneous spiking rate across the neurons within the microcolumn.  Biologically, this kind of learning coherency among neurons in a microcolumn would require particular sensitivities of different layers to signals coming from the superficial layer neurons, which are the only ones that have all the integrated information in one place to produce a suitable learning signal.  Alternatively, these other neurons may exhibit a different form of learning that produces a sufficient approximation to the computationally prescribed learning signal.  Further investigation into these issues is required.
\item How temporally crisp and coordinated across cortical areas does the layer 5b neuron firing need to be for the temporal context representation to work effectively?  For example, it is possible that different cortical columns can update at different time intervals, providing more of a overlapping, distributed temporal encoding.  This issue bears on the level of alpha power observed in different areas across different conditions, and how much alpha is synchronized across areas.  We will need to conduct various levels of computational simulations to test how robust the model is to these parameters, and indeed whether they might benefit overall function, in the same way that distributed overlapping neural representations are beneficial in many ways.
\item Relatedly, there is considerable evidence that the alpha rhythm can be entrained by external stimuli, which is important for ensuring that the temporal context and learning dynamics are properly organized around the actual flow of events in the environment.  In contrast, our LeabraTI model relies on a simple fixed discretization of time, uniform for all neurons in the network.  Thus, we will need to explore the mechanisms that support alpha phase entrainment, and incorporate these into our model.  This may require the development of a more biologically-detailed implementation of the model.
\item Does the time window of temporal integration change in different parts of the brain?  Empirically, alpha power tends to be more strongly associated with posterior neocortex, while the faster beta rhythm appears to be more prominent in frontal cortex, and the slower theta dominates in the hippocampus.  If we simplify the theta as 5Hz, then it would correspond to two alpha cycles --- this would suggest the appealing idea that episodic memories encoded in the hippocampus are built upon an integration over two underlying alpha cycles in posterior neocortex.  Meanwhile, the beta rhythm in frontal cortex could correspond to a 2x faster updating ability relative to the posterior cortical alpha rhythm, suggesting that these executive control signals could target minus or plus phases to drive learning effects, for example.  More detailed investigation of these possibilities is required, along with incorporation of such dynamics into larger-scale models with these interacting brain areas.
\end{itemize}

\subsection{Conclusions}

In conclusion, we have outlined a comprehensive and ambitious proposal for understanding how some detailed features of the neocortical system (including the thalamocortical loops) can support a computationally powerful predictive learning mechanism.  This proposal builds upon the solid foundation of work done with the simple recurrent network (SRN) framework at a more purely computational level.  Our proposal provides a novel way to achieve predictive learning in the neocortex, compared to other existing models, and thus provides an important point of comparison for deriving theoretically-motivated experiments to further understand how the neocortex supports all of its amazing functions.  

\section{Acknowledgments}

Supported by: ONR grant N00014-13-1-0067, ONR N00014-10-1-0177, ONR D00014-12-C-0638, and Supported by the Intelligence Advanced Research Projects Activity (IARPA) via Department of the Interior (DOI) contract number D10PC20021. The U.S. Government is authorized to reproduce and distribute reprints for Governmental purposes notwithstanding any copyright annotation thereon. The views and conclusions contained hereon are those of the authors and should not be interpreted as necessarily representing the official policies or endorsements, either expressed or implied, of IARPA, DOI, or the U.S. Government.

\clearpage

\clearpage
\section{Appendix: Computational Model Details}

This appendix provides more information about the object recognition model.  The purpose of this information is to give more detailed insight into the model's function beyond the level provided in the main text, but with a model of this complexity, the only way to really understand it is to explore the model itself.  It is available for download at \verb\http://grey.colorado.edu/CompCogNeuro/index.php/CCN_Repository\.  And the best way to understand this model is to understand the framework in which it is implemented, which is explained in great detail, with many running simulations explaining specific elements of functionality, at \verb\http://ccnbook.colorado.edu\.  

\subsection{Structure of the model}

This information is largely the same as for the model we are building upon \cite{OReillyWyatteHerdEtAl13}, but we have changed the scale of the model and several other features to incorporate the LeabraTI learning mechanisms.

\subsubsection{Early Visual Image Processing}

In the mammalian brain, the retina and LGN compress the visual input into an
efficient contrast-coded representation using center-surround contrast filters
that are radially symmetric (which can be nicely approximated by the
difference of two Gaussians, \nopcite{Enroth-CugellRobson66,Young87}). Then area
V1 encodes orientation and other features building upon this basic
contrast-enhanced input \cite{HubelWiesel62}.   We compress this chain of
filters into a single step by using oriented Gabor filters, which are defined
as a Gaussian-shaped spatial weighting multiplying a planar sine wave oriented
in a given direction:
\begin{equation}
 g(x,y) = e^{-\left(\frac{x^2}{2 \sigma_x^2} + \frac{y^2}{2 \sigma_y^2}\right)}
 \sin\left(\frac{2\pi y}{\lambda}\right)
 \label{eq:Gabor}
\end{equation}
where the sine wave moves along the y axis (corresponding to a horizontal
orientation tuning), and the Gaussian has differential width terms
($\sigma_x$, $\sigma_y$) for each axis. To obtain different orientations, the
coordinates x,y are rotated by a given angle $\theta$ relative to the original
coordinates of the filter ($x'$, $y'$):
\begin{eqnarray}
 x & = & x' \cos(\theta) - y' \sin(\theta) \nonumber \\
 y & = & y' \cos(\theta) + x' \sin(\theta)
 \label{eq:rotate}
\end{eqnarray}
The filter is always normalized to a zero sum in the discrete kernel that is
actually used, to ensure that a uniform illumination produces no activation.

In line with other established models of object recognition in cortex (e.g.,
\nopcite{WallisRolls97,RiesenhuberPoggio99,DaileyCottrell99,MasquelierThorpe07}), these filtering
operations provide a reasonable approximation to the coding properties of V1
simple cells.   The model processes each image at two different spatial frequencies (SF),
``high'' and ``medium'', each of which employs 4 orientations of tuning, times
2 for on vs. off-center polarity. The high-SF pathway uses Gabor filters
rendered on a 6x6 pixel kernel, with a wavelength $\lambda=6$, and Gaussian
width terms of $\sigma_x = 1.8$ and $\sigma_y = 1.2$. The medium-SF pathway
uses Gabor filters that are twice as large (12x12 kernel, wavelength
$\lambda=12$, and Gaussian width terms of $\sigma_x = 3.6$ and $\sigma_y =
2.4$). These filters are applied in a half-overlapping fashion to the image,
such that adjacent V1 simple units process spatial locations that are one half
wavelength ($\frac{\lambda}{2}$) away from their neighbors.

The input ``retina'' resolution is only 24x24 pixels, and the high frequency V1 simple filters are computed centered on each pixel (spacing = 1), producing a 24x24x8 (where 8 refers to 4 orientations x 2 polarities) dimensional output, while the medium frequency have a spacing of 2 and produce a 12x12x8 dimensional output.

The output of the V1 simple cells is computed using the FFFB inhibitory dynamics and
point-neuron activation function of Leabra (described below), applied to a net
input that is the positive-rectified (values less than 0 are clipped to 0)
result of convolving the Gabor kernel with the input image. There are two
levels of inhibitory competition -- the primary is within the group of
different orientation and polarity tunings for the same spatial location
(i.e., 8 units = 4 orientations x 2 polarities). This {\em unit group} level competition may
reflect competition at the level of the {\em hypercolumn} in the brain. This
inhibition is computed with a gi value of 2.0.  The secondary level of competition involves a
spread of the unit-group level competition across the entire layer of such
units, with a discounted gain multiplier and a MAX operation such that the
stronger of the unit-group or discounted layer-level competition holds (see
FFFB section of Leabra algorithm section for details).

\subsection{Structure of Higher Layers (Extrastriate, Inferotemporal, Output,
 Semantics)}

Proceeding from the V1 simple inputs at the two different spatial frequencies
(high and medium), the model captures the general response properties of
extrastriate cortex (V2, V3, V4) and inferotemporal (IT) cortex. As a purely
computational convenience in configuring the network, the model's V2
layers remain spatial-frequency specific (in the brain, we would expect these
to all be intermixed), which then merge into unitary V3, V4, and IT layers, and IT then
feeds into a naming output layer, and a semantics output layer (Figure
\ref{fig.objrec_ti_net}). All connections are bidirectional, except those from V2
back to V1 in the implicit prediction case -- for explicit prediction, V2 does project back to V1 to generate the minus phase expectation.  

All of the units are allowed to learn based on the Leabra learning mechanism
(described in the next section).  Once trained, the single model can
discriminate all trained object categories --- in contrast, other prevalent
feedforward models (e.g., \nopcite{RiesenhuberPoggio99,MasquelierThorpe07}) use
binary classifiers that would require \textit{N} classifiers to differentiate
among \textit{N} categories. Thus, the overall solution to the invariant
object recognition problem that the model develops is qualitatively similar
with the prevalent feedforward models, yet is also realizable using a
homogeneous, biologically plausible set of mechanisms.

Here are the detailed parameters for each layer in the network (note that
15-25\% activity levels is the default for Leabra models of the cortex, based
on biological estimates; \nopcite{OReillyMunakata00,OReillyMunakataFrankEtAl12}):
\begin{itemize}
\item {\bf V2}: 25 units per unit group / hypercolumn, arranged into
  a 12x12 topographical grid for the high spatial frequency
  layer, and a 6x6 grid for the medium spatial frequency
  layer. Each unit receives from a topographically-corresponding 4x4 grid of
  V1 unit groups, with 1/2 overlap among neighboring unit groups. FFFB inhibition produces roughly  10\% activity within each unit group.
\item {\bf V3:} 576 total units receiving a full projection from all V2 units (and projecting bidirectionally back to them), and from the V2\_sum layer that summarizes the V2 unit groups with a single unit. 
\item {\bf V4:} 64 units per unit group / hypercolumn, arranged into a 3x3 topographical grid, and receiving a retinotopically-organized projection from 4x4 V2h unit groups, and from 8x8 V2m unit groups, half overlapping as before.
\item {\bf IT}: 200 total units receiving a full projection from all of the V3 and V4
 units (and projecting bidirectionally back to them), with a 15\% kWTA
 activity level (no unit group sub-structure)
\item {\bf Naming Output}: 200 units receiving a full projection from the IT
  (and projecting completely back to it), with a kWTA activity level of 1\%.
  This localist (single active unit) representation of output names is a
  computational simplification, standing in for the full phonological
  production pathways.
\end{itemize}

\subsection{Model Algorithms}

The model was implemented using the Leabra framework, which is described in detail in \incite{OReillyMunakataFrankEtAl12}, \incite{OReillyMunakata00}, \incite{OReilly01}, and summarized here.  See Table~\ref{tab.sim_params} for a listing of parameter values, nearly all of which are at their default settings.  These same parameters and equations have been used to simulate over 40 different models in \incite{OReillyMunakataFrankEtAl12} and \incite{OReillyMunakata00}, and a number of other research models.  Thus, the model can be viewed as an instantiation of a systematic modeling framework using standardized mechanisms, instead of constructing new mechanisms for each model.  

This version of Leabra contains three primary differences from the original \cite{OReillyMunakata00}: the activation function is slightly different, in a way that allows units to more accurately reflect their graded excitatory input drive, the inhibition function is much simpler and more biologically realistic, and the learning rule takes a more continuous form involving contrasts between values integrated over different time frames (i.e., with different time constants), which also produces a combination of error-driven and self-organizing learning within the same simple mathematical framework.  These modifications are described in detail in an updated version of the \incite{OReillyMunakata00} textbook, in \incite{OReillyMunakataFrankEtAl12}.  This new learning algorithm goes by the acronym of XCAL (temporally eXtended Contrastive Attractor Learning), and it replaces the combination of Contrastive Hebbian Learning (CHL) and standard Hebbian learning used in the original Leabra framework.

\subsubsection{Pseudocode}

The pseudocode for Leabra is given here, showing exactly how the pieces of the algorithm described in more detail in the subsequent sections fit together. The individual steps are repeated for each event (trial), which can be broken down into a {\em minus} and {\em plus} phase, followed by a synaptic weight updating function. Generally speaking, the minus phase represents the system's expectation for a given input and the plus phase represents the observation of the outcome. The difference between these two phases is then used to compute the updating function that drives learning. Furthermore, each phase contains a period of {\em settling} (measured in {\em cycles}) during which the activation values of each unit are updated taking into account the previous state of the network. Some units are {\em clamped}, or have fixed activation values and are not subject to this updating rule (e.g., V1 input in the minus phase, V1 input and Output in the plus phase).

Outer loop: For each event (trial) in an epoch:
\begin{enumerate}
\item Iterate over minus and plus phases of settling for each event.
 \begin{enumerate}
 \item At start of settling, for all units:
  \begin{enumerate}
  \item Initialize all state variables (activation, $V_m$, etc).
  \item Clamp external patterns (V1 input in minus phase, V1 input \& Output in plus phase).
  \end{enumerate}
 \item During each cycle of settling, for all non-clamped units:
  \begin{enumerate}
  \item Compute excitatory netinput ($g_e(t)$ or $\eta_j$,
   eq~\ref{eq.net_in_avg}).
  \item Compute FFFB inhibition for each layer, based on average net input and activation levels within the layer (eq~\ref{eq.fffb})
  \item Compute point-neuron activation combining excitatory input and inhibition (eq~\ref{eq.vm}).
  \item Update time-averaged activation values (short, medium, long) for use in learning.
  \end{enumerate}
 \end{enumerate}
 \item After both phases update the weights, for all connections:
 \begin{enumerate}
 \item Compute XCAL learning as function of short, medium, and long time averages.
 \item Increment the weights according to net weight change.
 \end{enumerate}
\end{enumerate}

\subsubsection{Point Neuron Activation Function} 

\begin{table}
 \centering
 \begin{tabular}{ll|ll} \hline
Parameter & Value & Parameter & Value \\ \hline
$E_l$ & 0.30 & $\overline{g_l}$ & 0.10 \\
$E_i$ & 0.25 & $\overline{g_i}$ & 1.00 \\
$E_e$ & 1.00 & $\overline{g_e}$ & 1.00 \\
$V_{rest}$ & 0.30 & $\Theta$  & 0.50 \\
$\tau$ & .3 & $\gamma$ & 80 \\ \hline
 \end{tabular}
 \caption{\small Parameters for the simulation (see equations in text
  for explanations of parameters). All are standard default parameter values.}
 \label{tab.sim_params}
\end{table}

Leabra uses a {\em point neuron} activation function that models the electrophysiological properties of real neurons, while simplifying their geometry to a single point. This function is nearly as simple computationally as the standard sigmoidal activation function, but the more biologically-based implementation makes it considerably easier to model inhibitory competition, as described below. Further, using this function enables cognitive models to be more easily related to more physiologically detailed simulations, thereby facilitating bridge-building between biology and cognition. We use normalized units where the unit of time is 1 msec, the unit of electrical potential is 0.1 V (with an offset of -0.1 for membrane potentials and related terms, such that their normal range stays within the $[0, 1]$ normalized bounds), and the unit of current is $1.0x10^{-8}$.

The membrane potential $V_m$ is updated as a function of ionic conductances $g$ with reversal (driving) potentials $E$ as follows:
\begin{equation}
 \Delta V_m(t) = \tau \sum_c g_c(t) \overline{g_c} (E_c - V_m(t))
 \label{eq.vm}
\end{equation}
with 3 channels ($c$) corresponding to: $e$ excitatory input; $l$ leak current; and $i$ inhibitory input. Following electrophysiological convention, the overall conductance is decomposed into a time-varying component $g_c(t)$ computed as a function of the dynamic state of the network, and a constant $\overline{g_c}$ that controls the relative influence of the different conductances. The equilibrium potential can be written in a simplified form by setting the excitatory driving potential ($E_e$) to 1 and the leak and inhibitory driving potentials ($E_l$ and $E_i$) of 0:
\begin{equation}
 V_m^\infty = \frac{g_e \overline{g_e}} {g_e
  \overline{g_e} + g_l \overline{g_l} + g_i \overline{g_i}} 
\end{equation}
which shows that the neuron is computing a balance between excitation and the opposing forces of leak and inhibition. This equilibrium form of the equation can be understood in terms of a Bayesian decision making framework \cite{OReillyMunakata00}.

The excitatory net input/conductance $g_e(t)$ or $\eta_j$ is computed as the proportion of open excitatory channels as a function of sending activations times the weight values:
\begin{equation}
 \eta_j = g_e(t) = \langle x_i \wij \rangle = \oneo{n} \sum_i x_i \wij
 \label{eq.net_in_avg}
\end{equation}
The inhibitory conductance is computed via the kWTA function described in the next section, and leak is a constant.

In its discrete spiking mode, Leabra implements exactly the AdEx (adaptive exponential) model \cite{BretteGerstner05}, which has been found through various competitions to provide an excellent fit to the actual firing properties of cortical pyramidal neurons \cite{GerstnerNaud09}, while remaining simple and efficient to implement. However, we typically use a rate-code approximation to discrete firing, which produces smoother more deterministic activation dynamics, while capturing the overall firing rate behavior of the discrete spiking model.

We recently discovered that our previous strategy of computing a rate-code graded activation value directly from the membrane potential is problematic, because the mapping between $V_m$ and mean firing rate is not a one-to-one function in the AdEx model. Instead, we have found that a very accurate approximation to the discrete spiking rate can be obtained by comparing the excitatory net input directly with the effective computed amount of net input required to get the neuron firing over threshold ($g_e^{\Theta}$), where the threshold is indicated by $\Theta$:
\begin{equation}
g_e^{\Theta} = \frac{g_i \overline{g}_i (E_i - V_m^{\Theta}) +
 \overline{g}_l(E_l - V_m^{\Theta})} {\overline{g}_e (V_m^{\Theta} - E_e)}
\end{equation}
\begin{equation}
 y_j(t) \propto g_e(t) - g_e^{\Theta}
\end{equation}
where $y_j(t)$ is the firing rate output of the unit.

We continue to use the Noisy X-over-X-plus-1 (NXX1) function, which starts out with a nearly linear function, followed by a saturating nonlinearity:
\begin{equation}
 y_j(t) = \oneo{\left(1 + \oneo{\gamma [g_e(t) - g_e^{\Theta}]_+} \right)}
\end{equation}
where $\gamma$ is a gain parameter, and $[x]_+$ is a threshold function that returns 0 if $x<0$ and $x$ if $x>0$. Note that if it returns 0, we assume $y_j(t) = 0$, to avoid dividing by 0. As it is, this function has a very sharp threshold, which interferes with graded learning learning mechanisms (e.g., gradient descent). To produce a less discontinuous deterministic function with a softer threshold, the function is convolved with a Gaussian noise kernel ($\mu=0$, $\sigma=.005$), which reflects the intrinsic processing noise of biological neurons:
\begin{equation}
 y^*_j(x) = \int_{-\infty}^{\infty} \oneo{\sqrt{2 \pi} \sigma}
 e^{-z^2/(2 \sigma^2)} y_j(z-x) dz
 \label{eq.conv}
\end{equation}
where $x$ represents the $[g_e(t) - g_e^{\Theta}]_+$ value, and $y^*_j(x)$ is the noise-convolved activation for that value. In the simulation, this function is implemented using a numerical lookup table.

There is just one last problem with the equations as written above: They don't evolve over time in a graded fashion.  In contrast, the Vm value does evolve in a graded fashion by virtue of being iteratively computed, where it incrementally approaches the equilibrium value over a number of time steps of updating.  Instead the activation produced by the above equations goes directly to its equilibrium value very quickly, because it is calculated based on excitatory conductance and does not take into account the sluggishness with which changes in conductance lead to changes in membrane potentials (due to capacitance).

To introduce graded iterative dynamics into the activation function, we just use the activation value ($y^*(x)$) from the above equation as a ''driving force'' to an iterative temporally-extended update equation:
\begin{equation}
  y_j(t) = y_j(t-1) + dt_{vm} \left(y_j^*(t) - y_j(t-1) \right)
 \label{eq.y_iter}
\end{equation}
This causes the actual final rate code activation output at the current time $t$, $y(t)$ to iteratively approach the driving value given by $y^*(x)$, with the same time constant $dt_{vm}$ that is used in updating the membrane potential.  In practice this works extremely well, better than any prior activation function used with Leabra.

\subsubsection{FFFB Inhibition}

Leabra computes a layer-level inhibition conductance value based on a combination of feed-forward (FF) and feed-back (FB) dynamics.  This is an advance over the more explicit kWTA (k-Winners-Take-All) function that was used previously, though it achieves roughly the same overall kWTA behavior, with a much simpler, more efficient, and biologically plausible formulation.  The FF component is based directly on the average excitatory net input coming into the layer ($<\eta>$), and the FB component is based on the average activation of units within the layer ($<act>$).  Remarkably, fixed gain factors on each of these terms, together with simple time integration of the FB term to prevent oscillations, produces results that are overall comparable to the kWTA dynamics, except that the activations of units in the layer retain more of a proportional response to their overall level of excitatory drive, which is desirable in many cases.

FFFB is conceptually just the sum of the FF and FB components, each with their own ff and fb gain factors, with an overall gain factor (gi) applied to both:
\begin{equation}
  g_i = \mbox{gi} \left( \mbox{ff} [<\eta> - \mbox{ff0}]_+ + \mbox{fb} <act> \right)
  \label{eq.fffb}
\end{equation}
where $[ x ]_+$ indicates the positive part of whatever it contains -- anything negative truncates to zero.  It is important to have a small offset on the FF component, parameterized by ff0 which is typically .1 --- this delays the onset of inhibition and allows the neurons to get a little bit active first.  To minimize oscillations, the feedback component needs to be time integrated, with a fast time constant of .7 -- just a simple exponential approach to the driving fb inhibition value was used:
\begin{equation}
  fb_i(t) = fb_i(t-1) + dt \left(\ \mbox{fb} <act> - fb_i(t-1) \right)
  \label{eq.fbi}
\end{equation}
Typically ff is set to 1.0, fb is 0.5, and the overall gain (gi) is manipulated to achieve desired activity levels -- typically it is around 2.2 or so.

\subsubsection{XCAL Learning} 

The full treatment of the new XCAL version of learning in Leabra is presented in \incite{OReillyMunakataFrankEtAl12}, but the basic equations and a brief motivation for them are presented here.

In the original Leabra algorithm, learning was the sum of two terms: an error-driven component and a Hebbian self-organizing component. In the new XCAL formulation, the error-driven and self-organizing factors emerge out of a single learning rule, which was derived from a biologically detailed model of synaptic plasticity by Urakubo et al.~\cite{UrakuboHondaFroemkeEtAl08}, and is closely related to the Bienenstock, Cooper \& Munro (BCM) algorithm \cite{BienenstockCooperMunro82}. In BCM, a Hebbian-like sender-receiver activation product term is modulated by the extent to which the receiving unit is above or below a long-term running average activation value:
\begin{equation}
 \Delta_{bcm} \wij = xy (y - \langle y^2 \rangle)
 \label{eq:bcm}
\end{equation}
($x$ = sender activation, $y$ = receiver activation, and $\langle y^2 \rangle$ = long-term average of squared receiver activation). The long-term average value acts like a dynamic plasticity threshold, and causes less-active units to increase their weights, while more-active units tend to decrease theirs (i.e., a classic homeostatic function). This form of learning resembles Hebbian learning in several respects, but can learn higher-order statistics, whereas Hebbian learning is more constrained to extract low-order correlational statistics. Furthermore, the BCM model may provide a better account of various experimental data, such as monocular deprivation experiments \cite{CooperIntratorBlaisEtAl04}.

\begin{figure}[ht!]
 \centering\includegraphics[height=2in]{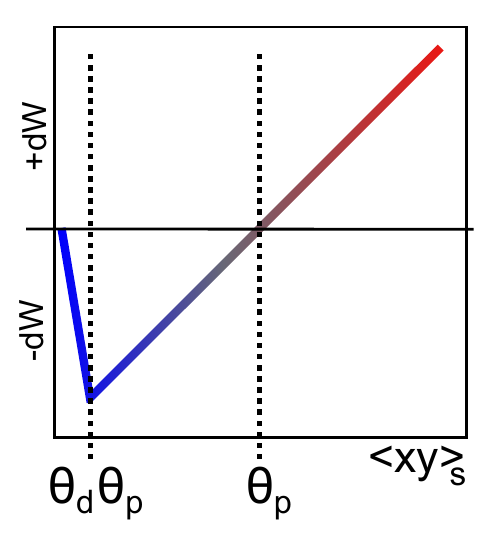}
 \caption{\small XCAL dWt function, shown with $\theta_p=0.5$, which determines
   the cross-over point between negative and positive weight changes, and
   $\theta_p \theta_d$ determines the inflection point at the left where the
   curve goes from a negative slope to a positive slope. This function fits
   the results of the highly detailed Urakubo et al \protect
   \cite{UrakuboHondaFroemkeEtAl08} model, with a correlation value of
   $r=0.89$.}
 \label{fig.xcal_dwt}
\end{figure}

The Leabra XCAL learning rule is based on a contrast between a sender-receiver activation product term (shown initially as just $xy$ -- relevant time scales of averaging for this term are elaborated below) and a dynamic plasticity threshold $\theta_p$ (also elaborated below), which are integrated in the XCAL learning function (Figure~\ref{fig.xcal_dwt}):
\begin{equation}
 \Delta_{xcal} \wij = f_{xcal} ( xy, \theta_p)
 \label{eq.xcal_simp}
\end{equation}
where the XCAL learning function was derived by fitting a piecewise-linear function to the Urakubo et al \cite{UrakuboHondaFroemkeEtAl08} simulation results based on synaptic drive levels (sender and receiver firing rates; the resulting fit was very good, with a correlation of $r=0.89$):
\begin{equation}
f_{xcal}(xy, \theta_p) = \left\{ \begin{array}{ll}
(xy - \theta_p) & \mbox{if} \; xy > \theta_p \theta_d \\
-xy (1 - \theta_d) / \theta_d & \mbox{otherwise} \end{array} \right.
\end{equation}
($\theta_d = .1$ is a constant that determines the point where the function reverses back toward zero within the weight decrease regime -- this reversalpoint occurs at $\theta_p \theta_d$, so that it adapts according to the dynamic $\theta_p$ value).

The BCM equation produces a curved quadratic function that has the same qualitative shape as the XCAL function (Figure~\ref{fig.xcal_dwt}). A critical feature of these functions is that they go to 0 as the synaptic activity goes to 0, which is in accord with available data, and that they exhibit a crossover point from LTD to LTP as a function of synaptic drive (which is represented biologically by intracellular Calcium levels). A nice advantage of the linear XCAL function is that, to first approximation, it is just computing the subtraction $xy - \theta_p$.

To achieve full error-driven learning within this XCAL framework, we just need to ensure that the core subtraction represents an error-driven learning term.  In the original Leabra, error-driven learning via the Contrastive Hebbian Learning algorithm (CHL) was computed as:
\begin{equation}
 \Delta_{chl} = x^+ y^+ - x^- y^-
 \label{eq:chl}
\end{equation}
where the superscripts represent the plus ($+$) and minus ($-$) phases. This equation was shown to compute the same error gradient as the backpropagation algorithm, subject to symmetry and a 2nd-order numerical integration technique known as the midpoint method, based the generalized recirculation algorithm (GeneRec; \cite{OReilly96}). In XCAL, we replace these values with time-averaged activations computed over different time scales:
\begin{itemize}
\item {\bf s} = short time scale, reflecting the most recent state of neural  activity (e.g., past 100-200 msec). This is considered the ``plus phase'' -- it represents the {\em outcome} information on the current trial, and in  general should be more correct than the medium time scale.
\item {\bf m} = medium time scale, which integrates over an entire  psychological ``trial'' of roughly a second or so -- this value contains a mixture of the ``minus phase'' and the ``plus phase'', but in contrasting it  with the short value, it plays the role of the minus phase value, or expectation about what the system thought should have happened on the  current trial.
\item {\bf l} = long time scale, which integrates over hours to days of processing -- this is the BCM-like threshold term.
\end{itemize}

Thus, the error-driven aspect of XCAL learning is driven essentially by the following term: 
\begin{equation}
 \Delta_{xcal-err} \wij = f_{xcal} ( x_s y_s, x_m y_m )
 \label{eq.xcal-err}
\end{equation}
However, consider the case where either of the short term values ($x_s$ or $y_s$) is 0, while both of the medium-term values are $>0$ -- from an error-driven learning perspective, this should result in a significant weight decrease, but because the XCAL function goes back to 0 when the input drive term is 0, the result is no weight change at all. To remedy this situation, we assume that the short-term value actually retains a small trace of the medium-term value:
\begin{equation}
 \Delta_{xcal-err} \wij = f_{xcal} ( \kappa x_s y_s + (1-\kappa) x_m y_m, x_m y_m)
 \label{eq.xcal-err2}
\end{equation}
(where $\kappa = .9$, such that only .1 of the medium-term averages are incorporated into the effective short-term average).

The self-organizing aspect of XCAL is driven by comparing this same synaptic drive term to a longer-term average, as in the BCM algorithm:
\begin{equation}
 \Delta_{xcal-so} \wij = f_{xcal} ( \kappa x_s y_s + (1-\kappa) x_m y_m, \gamma_l y_l)
 \label{eq.xcal-selforg}
\end{equation}
where $\gamma_l = 3$ is a constant that scales the long-term average threshold term (due to sparse activation levels, these long-term averages tend to be rather low, so the larger gain multiplier is necessary to make this term relevant whenever the units actually are active and adapting their weights).

Combining both of these forms of learning in the full XCAL learning rule amounts to computing an aggregate $\theta_p$ threshold that reflects a combination of both the self-organizing long-term average, and the medium-term minus-phase like average:
\begin{equation}
 \Delta_{xcal} \wij = f_{xcal} ( \kappa x_s y_s + (1-\kappa) x_m y_m, \lambda
 \gamma y_l + (1-\lambda) x_m y_m)
 \label{eq.xcal}
\end{equation}
where $\lambda = .01$ is a weighting factor determining the mixture of self-organizing and error-driven learning influences (as was the case with standard Leabra, the balance of error-driven and self-organizing is heavily weighted toward error driven, because error-gradients are often quite weak in comparison with local statistical information that the self-organizing system encodes).

The weight changes are subject to a soft-weight bounding to keep within the $0-1$ range:
\begin{equation}
 \Delta_{sb} \wij = [\Delta_{xcal}]_+ (1-\wij) + [\Delta_{xcal}]_- \wij
 \label{eq.err_soft_bound}
\end{equation}
where the $[]_+$ and $[]_-$ operators extract positive values or negative-values (respectively), otherwise 0.

Finally, as in the original Leabra model, the weights are subject to contrast enhancement, which magnifies the stronger weights and shrinks the smaller ones in a parametric, continuous fashion. This contrast enhancement is achieved by passing the linear weight values computed by the learning rule through a sigmoidal nonlinearity of the following form:
\begin{equation}
 \hat{w}_{ij} = \oneo{1 + \left(\frac{\wij}{\theta (1-\wij)}\right)^{-\gamma}}
 \label{eq.wt_off}
\end{equation}
where $\hat{w}_{ij}$ is the contrast-enhanced weight value, and the sigmoidal function is parameterized by an offset $\theta$ and a gain $\gamma$ (standard defaults of 1 and 6, respectively, used here). 

\subsubsection{TI Context} 

At the end of every plus phase, a new TI context net input is computed from the dot product of the context weights times the sending activations, just as in the standard net input:
\begin{equation}
 \eta_{ti} = \langle x_i \wij \rangle = \oneo{n} \sum_i x_i \wij
 \label{eq.net_in_ti}
\end{equation}
This net input is then added in with the standard net input (equation~\ref{eq.net_in_avg}) at each cycle of processing.

Learning of the context weights occurs through the superficial neuron's error signal, as discussed in the main text, with the sending activation being the {\em prior} time step's plus phase activation.  We use a simple delta rule given that the context representations are static throughout the trial.
\begin{equation}
 \Delta_{ti} \wij = x_{t-1} \left( y^+_j - y^-_j \right)
 \label{eq.ti_delta}
\end{equation}
where $y$ is the superficial neural activation in the plus and minus phases, as denoted, and $x_{t-1}$ is the sending activation from the prior plus phase.

In general, these context projections exist for all standard projection pathways in the model, in addition to the self-context of the layer onto itself.

\end{document}